\documentclass[preprint,12pt]{elsarticle}

\usepackage{latexsym}
\usepackage{amssymb}
\usepackage{amsmath}
\usepackage{mathrsfs}

\journal{Journal of Geometry and Physics}

\begin{document}

\begin{frontmatter}



\title{Symmetries, integrals and hierarchies of new (3+1)-dimensional bi-Hamiltonian systems of Monge--Amp\`ere type}


\author[bu]{M. B. Sheftel}
\ead{mikhail.sheftel@boun.edu.tr}

\author[ytu]{D. Yaz{\i}c{\i}}
\ead{yazici@yildiz.edu.tr}

\address[bu]{Department of Physics, Bo\u{g}azi\c{c}i University, Bebek, 34342 Istanbul, Turkey}
\address[ytu]{Department of Physics, Y{\i}ld{\i}z Technical University, Esenler, 34220 Istanbul, Turkey}

\begin{abstract}
We study point symmetries, the corresponding conserved densities and hierarchies of four new bi-Hamiltonian heavenly systems in 3+1 dimensions which we discovered recently. We exhibit an important role played by the inverse recursion operators in the description of the hierarchies. Their use is twofold, either to provide the correct bi-Hamiltonian representation or to generate nonlocal symmetry flows. Invariant solutions w.r.t. nonlocal symmetries will generate (anti-)self-dual gravitational metrics which do not admit Killing vectors which is a characteristic feature of $K3$ gravitational instanton.
\end{abstract}


\end{frontmatter}

% \linenumbers

\section{Introduction}
\label{intro}

In our previous paper \cite{S_Y} we considered (3+1)-dimensional second-order evolutionary PDEs
\[f(u_{ij}) - u_{tt}g(u_{ij}) = 0\]
where the unknown $u=u(t,\{z_i\})$ enters only in the form of the 2nd-order partial derivatives $u_{ij}$, $u_{ti}$ ($i,j=1,2,3$) and there is no explicit dependence on independent variables.
We have proved that all such equations, which possess a Lagrangian, have the Monge--Amp\`ere form which is defined as a linear relation among all possible minors of the Hessian matrix of $u$. In a two-component form all these equations become Hamiltonian systems. Using our approach of ``skew-factorization'' of the determining equation for symmetries as a tool for producing recursion operators, we discovered four nonequivalent new bi-Hamiltonian systems integrable in the sense of Magri \cite{magri}. The method for finding the recursion operators in \cite{S_Y} extends the method of A. Sergyeyev from \cite{Artur}. An invariant differential-geometric characterization of the Monge-Amp\`ere property is given in \cite{DFKN}.

Our interest to this class of equations is caused by the fact that all known heavenly equations governing (anti-)self-dual gravity belong to this class \cite{pleb,nns,nsky,sy,SYM,ferdub}. We are interested in gravitational instanton solutions \cite{GibHawk}, and most of all in the famous $K3$ instanton \cite{yau} which explicit gravitational metric is still unknown. One of the main properties of $K3$ is that it does not admit Killing vectors, i.e. continuous symmetries. For this property to be satisfied we need solutions of heavenly equations that are noninvariant w.r.t. any point symmetries to avoid symmetry reduction in a number of independent variables in the solutions. Our tool for constructing such solutions is their invariance w.r.t. nonlocal symmetries which does not require symmetry reduction. We must point out that there is a fairly extensive literature on solutions invariant under nonlocal symmetries, e.g. \cite{Bluman} and references therein.

In a subsequent publication we will obtain (anti-)self-dual gravitational metrics which are governed by our new heavenly bi-Hamiltonian systems using the methods which we applied earlier to the general heavenly equation and modified heavenly equation \cite{mash,shya}.

In this paper we study the new bi-Hamiltonian systems in more detail. We determine point symmetries and associated conservation laws (integrals of motion) for these systems together with an algorithm of reconstructing the integral generating a given symmetry. We utilize the Magri integrability by studying hierarchies of the bi-Hamiltonian systems. The objects of special interest to us are nonlocal symmetry flows. Due to the nonlocality, invariant (stationary) solutions of nonlocal flows will not experience symmetry reductions in the number of independent variables and generate gravitational metrics without Killing vectors.

The paper is organized as follows.
In Section \ref{sec-sym}, we gather results on point symmetries of our new bi-Hamiltonian systems which we call system $I$, system $II$, system $III$ and system $IV$. In subsections \ref{I.1}, \ref{II.1}, \ref{III.1} and \ref{IV.1} we present point symmetries and their commutator algebras for systems $I$, $II$, $III$ and $IV$, respectively. In Section \ref{sec-int}, we present an algorithm for determining conserved densities which generate known symmetries via the Hamiltonian structure and apply it in subsections \ref{I.2}, \ref{II.2}, \ref{III.2} and \ref{IV.2} to obtain the conserved densities for systems $I$, $II$, $III$ and $IV$, respectively. In Section \ref{sec-hier}, we review some basic properties of hierarchies of bi-Hamiltonian systems. A new feature is the utilization of the inverse recursion operators which allow us to move along the hierarchy chain not only in the right direction but also in the left direction. In subsections \ref{I.3}, \ref{II.3}, \ref{III.3} and \ref{IV.3} we describe in detail the hierarchies of systems $I$, $II$, $III$ and $IV$, respectively. The most important for us are the nonlocal flows in each hierarchy because their stationary (invariant) solutions need not to experience symmetry reduction and hence the corresponding gravitational metrics will not admit Killing vectors, which is a characteristic property of the $K3$ instanton.

\section{Point symmetries}
\setcounter{equation}{0}
\label{sec-sym}

In this section we study point symmetries and corresponding integrals of each of our new bi-Hamiltonian systems (7.3), (9.11), (9.22) and (9.33) from our preceding paper \cite{S_Y}, which we call now as systems $I$, $II$, $III$ and $IV$, respectively.
We skip system (9.2) from \cite{S_Y} because it can be obtained from system I by a permutation of indices combined with an appropriate permutation of coefficients.

For the sake of compactness, in the following we utilize the operators $L_{ij(k)} = u_{jk}D_i - u_{ik}D_j$ where $D_i$ denotes the total derivative with respect to $z_i$.

\subsection{System I}
\label{I.1}
\setcounter{equation}{0}

System $I$ reads
\begin{eqnarray}
&&u_t = v,\quad v_t = \frac{1}{u_{23}}\{v_2v_3 - c_4L_{12(3)}[v] - c_5L_{23(2)}[v] - c_8L_{23(1)}[v]\nonumber\\
&& \mbox{} - c_9L_{12(3)}[u_1] - c_{10}L_{23(2)}[u_1]\} = q
 \label{6.5}
\end{eqnarray}
 with the  condition $c_{10}=c_5c_9/c_8$.

Generators of point symmetries have the form ($\partial_i = \partial_{z_i}$)
\begin{eqnarray}
&& X_1 = \partial_1,\quad X_2 = u\partial_u + v\partial_v,\quad X_3 = \partial_t\nonumber\\
&& X_4 = t\partial_t + z_1\partial_1 + z_2\partial_2 + u\partial_u,\quad X_a = a(z_3)\partial_u,\quad Y_b = b(z_3)\partial_3\nonumber\\
&& X_\infty = c(\zeta)\partial_2 + \bigl(A(\omega_+) + B(\omega_-) + e(\zeta)\bigr)\partial_u
\label{sym6.5}\\
&&\mbox{} + \left\{\left(c_4+\sqrt{c_4^2-4c_9}\right)A'(\omega_+) + \left(c_4-\sqrt{c_4^2-4c_9}\right)B'(\omega_-)\right\}\partial_v\nonumber
\end{eqnarray}
where $\zeta=c_5z_1-c_8z_2$, $\omega_\pm = \left(c_4\pm\sqrt{c_4^2-4c_9}\right)t-2z_1$.

The corresponding two-component symmetry characteristics read
\begin{eqnarray}
&&\varphi_1 = - u_1,\;\psi_1 = - v_1,\quad \varphi_2=u,\;\psi_2 = v,\nonumber\\
&&\varphi_3 = - v,\;\psi_3= - q,\quad \varphi_4=u-tv-z_1u_1-z_2u_2-z_3u_3\nonumber\\
&&\psi_4=-t q-z_1v_1-z_2v_2-z_3v_3,\quad \varphi_a=a(z_3),\;\psi_a=0,\quad\varphi_b=-b(z_3)u_3\nonumber\\
&&\psi_b=-b(z_3)v_3,\quad\varphi_\infty=A(\omega_+)+B(\omega_-)+e(\zeta) -c(\zeta)u_2
 \label{char6.5}\\
&&\psi_\infty=\left(c_4+\sqrt{c_4^2-4c_9}\right)A'(\omega_+) + \left(c_4-\sqrt{c_4^2-4c_9}\right)B'(\omega_-)-c(\zeta)v_2.\nonumber
\end{eqnarray}

For simplicity, we will determine Hamiltonian density for the flow generated by $X_\infty$ in \eqref{sym6.5} only if it commutes with $\partial_t$
and so has no $t$-dependence
\begin{equation}
  X_\infty =  X_{(c,e)} = c(\zeta)\partial_2 + e(\zeta)\partial_u
\label{XinfI }
\end{equation}
with the characteristic
\begin{equation}
    \varphi_{(c,e)} = e(\zeta) - c(\zeta)u_2,\quad \psi_{(c,e)} = - c(\zeta)v_2.
 \label{charinfI}
\end{equation}

In the following tables of commutators of point symmetries the commutator $[X_i,X_j]$ stands at the intersection of $i$th row and $j$th column,
$\tilde c = \zeta c' - c$, $\tilde e = \zeta e' - e$, $\hat c = \sigma c' - c\sigma'$, $\hat e = \varepsilon e' - e\varepsilon'$ and similarly
for $\tilde\sigma$ and $\tilde\varepsilon$.
\begin{table}[ht]
%\hspace{-24pt}
\begin{tabular}{|c|c|c|c|c|c|c|c|}
\hline    &$X_1$&$X_2$  &$X_3$& $X_4$ &$X_a$ &$Y_b$ &$X_{(c,e)}$
\\ \hline
    $X_1$ & $0$ & $0$& $0$ & $X_1$ &$0$  &$0$   &$c_5X_{(c',e')}$
\\ \hline
    $X_2$ &$0$& $0$ & $0$   & $0$ &$-X_a$&$0$& $-X_{(0,e)}$
\\ \hline
    $X_3$ &$0$  &$0$  & $ 0$  &$X_3$  &$0$& $0$& $0$
\\ \hline
    $X_4$ &$-X_1$& $0$& $-X_3$  & $0$& $-X_a$& $0$& $X_{(\tilde c,\tilde e)}$
\\ \hline
    $X_a$ &$0$& $X_a$& $0$& $X_a$& $0$& $0$& $0$
\\ \hline
    $Y_b$ &$0$& $0$& $0$& $0$& $0$& $0$& $0$
\\ \hline
$X_{(\sigma,\varepsilon)}$ &$-c_5X_{(\sigma',\varepsilon')}$& $X_{(0,\varepsilon)}$& $0$& $-X_{(\tilde \sigma,\tilde \varepsilon)}$& $0$& $0$& $X_{(\hat c,\hat e)}$
\\ \hline
\end{tabular}
\caption{Commutators of point symmetries of the system I.}
\end{table}
\newpage

\subsection{System II}
\label{II.1}
\setcounter{equation}{0}

System $II$ has the form
\begin{equation}
 u_t = v,\quad v_t =  q = \frac{1}{\Delta}\bigl\{v_2\bigl(\hat\Delta[v] - \hat c[u_3]\bigr) + v_3\hat c[u_2]\bigr\}
 \label{6.9}
\end{equation}
where we use the notation from our paper \cite{S_Y}
\begin{equation}
 \hat{\Delta} = a_8D_1 + a_{10}D_2 + a_{11}D_3,\quad \Delta = \hat\Delta[u_2], \quad \hat c = c_8D_1 + c_7D_3 .
 \label{not}
\end{equation}

Generators of point symmetries have the form
\begin{eqnarray}
&& X_1 = u\partial_u + v\partial_v,\quad X_a = a(z_1)\left\{(a_8z_3+c_8t)\partial_u + c_8\partial_v\right\}\nonumber\\
&& Y_b = - b_v(z_1,v)\partial_t + (b-vb_v)\partial_u \nonumber\\
&& X_\infty = \left\{c_8(a_8z_3 + c_8t)c'(z_1) + a_8\bigl(\delta E(\zeta) - c_7c(z_1)\bigr)\right\}\partial_t + c_8c(z_1)\partial_1 \nonumber\\
&& \mbox{} + \left\{g(\sigma) - a_{10}c_8^2E(\zeta)\right\}\partial_2 - c_8\bigl(E(\zeta) - c_7c(z_1)\bigr)\partial_3 \nonumber\\
&& \mbox{} + \left(\Phi(\sigma) - d(\zeta)\right)\partial_u - c_8^2 c'(z_1)v\partial_v.
\label{sym6.9}
\end{eqnarray}
Here $\zeta=c_7z_1 - c_8z_3$, $\sigma=a_{10}\zeta + \delta z_2$, $\delta = a_{11}c_8-a_8c_7$.

We will determine Hamiltonian density for the flow generated by $X_\infty$ in \eqref{sym6.9} only if it commutes with $\partial_t$: $[\partial_t,X_\infty] = c_8^2c'(z_1)\partial_t = 0$. Hence c is constant and $X_\infty$ becomes
\begin{equation}
  \hat X = a_8\delta e(\zeta)\partial_t + cc_8^2\partial_1 + \hat G(\sigma,\zeta)\partial_2 - c_8\delta e(\zeta)\partial_3 + \Phi(\sigma)\partial_u \equiv
\hat X_{(c,e,\hat G,\hat\Phi)}
\label{XinfII}
\end{equation}
where $e(\zeta) = E(\zeta) - cc_7/\delta$, $\hat G(\sigma,\zeta) = g(\sigma) - a_{10}c_8^2c_7c/\delta - a_{10}c_8^2e(\zeta)$, $\hat\Phi(\sigma,\zeta) = \Phi(\sigma) - d(\zeta)$ are new arbitrary functions of $\sigma$ and $\zeta$.

We note that some obvious symmetries are obtained as particular cases of the generators \eqref{sym6.9} and \eqref{XinfII}, e.g. $Y_{b=-v}=\partial_t$,
$\hat X_{(c=1,e=\hat G=\hat\Phi=0)} = c_8^2\partial_1$, $\hat X_{(c=e=\hat\Phi=0,\hat G=1)} = \partial_2$, $\hat X_{(e=1,c=\hat G=\hat\Phi=0)} = \delta(a_8\partial_t - c_8\partial_3)$ which is effectively\\[2pt] $\partial_3$ because we have already obtained the symmetry $\partial_t$.

Two-component symmetry characteristics of the generators \eqref{sym6.9} and \eqref{XinfII} read
\begin{eqnarray}
&&\varphi_1 = u,\;\psi_1 = v,\quad \varphi_a= a(z_1)(a_8z_3 + c_8t),\;\psi_a = c_8a(z_1),\; \nonumber\\
&&\varphi_b = b(z_1,v),\quad  \psi_b = b_v q,\quad \hat\varphi = \hat\Phi(\sigma,\zeta) - a_8\delta e(\zeta)v - c_8^2c u_1 \nonumber\\
&&\mbox{} - \hat G(\sigma,\zeta)u_2 + c_8\delta e(\zeta)u_3,\quad \hat\psi = - \frac{a_8\delta e(\zeta)}{\Delta}\bigl\{v_2\bigl(\hat\Delta[v] - \hat c[u_3])
 + v_3\hat c[u_2] \bigr\} \nonumber\\
&&\mbox{} - c_8^2cv_1 - \hat G(\sigma,\zeta)v_2 + c_8\delta e(\zeta)v_3.
 \label{char6.9}
\end{eqnarray}

The table of commutators of point symmetries generators of the system $II$ has the form

\begin{table}[ht]
\begin{center}
\begin{tabular}{|c|c|c|c|c|}
\hline    &$X_1$&$X_a$  &$Y_b$& $\hat X$
\\ \hline
    $X_1$ & $0$ & $-X_a$& $Y_{vb_v-b}$ & $-\hat X_{(c=e=\hat G=0)}$
\\ \hline
    $X_a$ &$X_a$& $0$ & $c_8aY_{b_v}$   & $-cc_8^2X_{a'(z_1)}$
\\ \hline
    $Y_b$ &$-Y_{vb_v-b}$  &$-c_8aY_{b_v}$  & $ 0$  &$-cc_8^2Y_{b_{z_1}}$
\\ \hline
    $\hat X$ &$\hat X_{(c=e=\hat G=0)}$& $cc_8^2X_{a'(z_1)}$& $cc_8^2Y_{b_{z_1}}$  & 0
\\ \hline
\end{tabular}
\end{center}
\caption{Commutators of point symmetries of the system II.}
\end{table}

\subsection{System III}
\label{III.1}
\setcounter{equation}{0}

System $III$ reads
\begin{eqnarray}
&& u_t = v,\quad v_t =  q = \frac{1}{u_{33}}\bigl\{v_3^2 - c_5(v_2u_{23} - v_3u_{22}) - c_6(v_1u_{33} - v_3u_{13})\nonumber\\
&&\mbox{} - c_7(v_2u_{33} - v_3u_{23}) - c_8(v_2u_{13} - v_3u_{12})\bigr\}. \nonumber\\
 \label{6.11}
\end{eqnarray}

Generators of point symmetries have the form
\begin{eqnarray}
&& X_1 = u\partial_u + v\partial_v,\quad X_\infty = -\bigl(f_v(\rho,v) - b(z_1)\bigr)\partial_t + c_6b(z_1)\partial_1\nonumber\\
&&\mbox{} + \frac{1}{c_8} \bigl(E(\zeta)+c_5c_6b(z_1)\bigr)\partial_2 + \left\{\Omega(\zeta) - \frac{\delta}{c_8^2}\bigl(E'(\zeta)z_1+c_6b(z_1)\bigr)\right. \nonumber\\
&&\left.\mbox{} - z_3E'(\zeta)\right\}\!\partial_3 + \left\{\left(\frac{c_8z_3}{\delta} + \frac{z_1}{c_8}\right)\chi(\zeta) + \alpha(\zeta)
- \frac{1}{c_8^2}A(z_1)\right.\nonumber\\
&&\left.\mbox{} + f(\rho,v) - vf_v\right\}\partial_u - c_6f_\rho(\rho,v)\partial_v
  \label{sym6.11}
\end{eqnarray}
where $\rho=z_1-c_6t$, $\zeta=c_5z_1 - c_8z_2$, $\delta = c_5c_6-c_7c_8$. We impose again the condition $[\partial_t,X_\infty]=0$ which implies that $f=f(v)$ is independent of $\rho$ in \eqref{sym6.11}. We denote
$\hat X = X_\infty|_{f=f(v)}$ which after appropriate redefinitions of arbitrary functions in \eqref{sym6.11} becomes
\begin{eqnarray}
&&\hat X = \bigl(c_8^2b(z_1) - f'(v)\bigr)\partial_t + c_6c_8^2b(z_1)\partial_1 + c_8\bigl(E(\zeta)+c_5c_6b(z_1)\bigr)\partial_2 \nonumber\\
&&\mbox{} - \left\{\omega(\zeta) + c_6b(z_1) + (\delta z_1 + c_8^2z_3)E'(\zeta)\right\}\!\partial_3 \nonumber\\
&&\mbox{}  + \left\{(\delta z_1 + c_8^2z_3)d(\zeta) + g(\zeta) - a(z_1) + f(v) - v f'(v)\right\}\partial_u
  \label{hatX}
\end{eqnarray}

The two-component characteristics of symmetries $X_1$ in \eqref{sym6.11}  and $\hat X$ in \eqref{hatX} read
\begin{eqnarray}
&&\varphi_1=u,\quad \psi_1=v\nonumber\\
&&\hat\varphi = f(v) - c_8^2b(z_1)v + (\delta z_1 + c_8^2z_3)d(\zeta) + g(\zeta) - a(z_1) - c_6c_8^2b(z_1)u_1 \nonumber\\
&&\mbox{} - c_8\bigl(E(\zeta) + c_5c_6b(z_1)\bigr)u_2 + \bigl\{(\delta z_1 + c_8^2z_3)E'(\zeta) + \delta c_6b(z_1) + \omega(\zeta)\bigr\}u_3 \nonumber\\
\label{char6.11}\\
&& \hat\psi = \bigl(f'(v) - c_8^2b(z_1)\bigr)q - c_6c_8^2b(z_1)v_1 - c_8\bigl(E(\zeta) + c_5c_6b(z_1)\bigr)v_2  \nonumber\\
&&\mbox{} + \bigl\{(\delta z_1 + c_8^2z_3)E'(\zeta) + \delta c_6b(z_1) + \omega(\zeta)\bigr\}v_3 \nonumber
\end{eqnarray}
where $q$ is defined in \eqref{6.11}.

\subsection{System IV}
\label{IV.1}
\setcounter{equation}{0}

System $IV$ reads
\begin{eqnarray}
&&u_t = v,\quad v_t = q \nonumber\\
&& q = \frac{1}{a_7u_{11}+a_8u_{12}+a_9u_{13}}\bigl\{(a_7v_1+a_8v_2+a_9v_3)v_1 - c_1(v_1u_{12}-v_2u_{11})\nonumber\\
&&\mbox{} - c_3(v_1u_{22}-v_2u_{12}) - c_4(v_1u_{23}-v_2u_{13})\bigr\}
 \label{6.13}
\end{eqnarray}

Generators of point symmetries have the form
\begin{eqnarray}
&& X_1 = u\partial_u + v\partial_v,\quad Y_a = - a_v(z_3,v)\partial_t + (a-va_v)\partial_u \nonumber\\
&& X_\infty = \left\{c_4^2c'(z_3)t + a_9\bigl[c_4z_2c'(z_3) - c_3c(z_3) + \alpha E(\zeta)\bigr]\right\}\!\partial_t \nonumber\\
&&\mbox{} + \left[c_1c_4c(z_3) - c_4\beta E(\zeta) + G(\sigma)\right]\partial_1 + c_4\bigl(c_3c(z_3) - \alpha E(\zeta)\bigr)\partial_2 \nonumber\\
&&\mbox{} + c_4^2c(z_3)\partial_3 + \left\{c_4(c_4t + a_9z_2)b^{\,\prime}(z_3) - a_9c_3 b(z_3) - \omega(\zeta) + F(\sigma)\right\}\partial_u\nonumber\\
&&\mbox{} + c_4\bigl(b^{\,\prime}(z_3) - c'(z_3)v\bigr)\partial_v
  \label{sym6.13}
\end{eqnarray}
where $\zeta=c_4z_2-c_3z_3$, $\sigma = \alpha z_1 - \beta z_2 - \gamma z_3$, $\alpha = a_8c_4 - a_9c_3$, $\beta = a_7c_4 - a_9c_1$,
$\gamma = a_8c_1 - a_7c_3$.

The condition $[\partial_t,X_\infty] = 0$ becomes
$c_4^2(c^{\,\prime}(z_3)\partial_t + b^{\,\prime}(z_3)\partial_u) = 0$ so that $b$ and $c$ are constants and
$X_\infty = \hat X + c\left(-a_9c_3\partial_t + c_4(c_3\partial_2 + c_4\partial_3)\right)$
where
\begin{eqnarray}
 &&\hat X = \alpha E(\zeta)(a_9\partial_t - c_4\partial_2) + \left(G(\sigma) - c_4\beta E(\zeta)\right)\partial_1  \nonumber\\
&&\mbox{} + \left[F(\sigma) - \Omega(\zeta)\right]\partial_u .
  \label{hatsymIV}
\end{eqnarray}
We note that we have the obvious translational symmetries $\partial_t$ and $\partial_{z_i}$ for $i=1,2,3$ as particular cases of the (combinations of) symmetries $Y_a$ and $X_\infty$ and hence we can skip the obvious symmetry $c\left(-a_9c_3\partial_t + c_4(c_3\partial_2 + c_4\partial_3)\right)$ and consider $\hat X$ instead of $X_\infty$.

\begin{table}[ht!]
\begin{center}
\begin{tabular}{|c|c|c|c|}
\hline    &$X_1$&$Y_a$  & $\hat X$
\\ \hline
    $X_1$ & $0$ & $Y_{va_v-a}$& $-\hat X_{(E=G=0)}$
\\ \hline
    $Y_a$ &$-Y_{va_v-a}$& $0$   & $0$
\\ \hline
    $\hat X$ &$\hat X_{(E=G=0)}$& $0$  & $0$
\\ \hline
\end{tabular}
\end{center}
\caption{Commutators of point symmetries of the system IV.}
\end{table}

Two-component symmetry characteristics read
\begin{eqnarray}
&& \varphi_1 = u,\quad\psi_1 = v,\qquad \varphi_a = a(z_3,v),\quad\psi_a = a_v(z_3,v)q \nonumber\\
&& \hat\varphi = F(\sigma) - \Omega(\zeta) - a_9\alpha E(\zeta)v - \left(G(\sigma) - c_4\beta E(\zeta)\right)u_1 + c_4\alpha E(\zeta)u_2 \nonumber\\
&& \hat\psi =  - a_9\alpha E(\zeta)q - \left(G(\sigma) - c_4\beta E(\zeta)\right)v_1 + c_4\alpha E(\zeta)v_2
 \label{char6.13}
\end{eqnarray}
where $q$ is defined in \eqref{6.13}.

\section{Conserved densities}
\setcounter{equation}{0}
\label{sec-int}

All the systems considered in \cite{S_Y} were shown to have the Hamiltonian form
\begin{equation}
  \left(\begin{array}{c} \displaystyle
  u_t\\ \displaystyle v_t
  \end{array}
  \right) = J_0 \left(\begin{array}{c}
 \delta_u H_1 \\ \delta_v H_1
  \end{array}
  \right)
\label{sysHam1}
\end{equation}
where $J_0$ is the Hamiltonian operator determining the structure of the Poisson bracket, $\delta_u, \delta_v$ are Euler-Lagrange operators \cite{olv} and $H_1$ is the Hamiltonian density.
The Hamiltonian structure provides a link between characteristics of symmetries and integrals of motion conserved by the Hamiltonian flows \eqref{sysHam1}.
Replacing time $t$ by the group parameter $\tau$ in (\ref{sysHam1}) and using $u_\tau = \varphi,\; v_\tau = \psi$ for symmetries in the evolutionary form,
we arrive at the Hamiltonian form of the Noether theorem for any conserved density $H$ of an integral of motion
\begin{equation}
  \left(\begin{array}{c} \displaystyle \varphi
  \\ \displaystyle \psi
  \end{array}
  \right) = J_0 \left(\begin{array}{c}
  \delta_u H \\ \delta_v H  \end{array}
  \right).
\label{Noether}
\end{equation}
To determine a conserved density $H$ that corresponds to a known symmetry with the characteristic $(\varphi, \psi)$ we use the inverse Noether theorem
\begin{equation}
 \left(\begin{array}{c}
  \delta_u H \\ \delta_v H  \end{array}
  \right) = K
  \left(\begin{array}{c} \varphi
  \\ \psi
  \end{array}
  \right)
\label{InvNoeth}
\end{equation}
where the symplectic operator $K = J_0^{-1}$ inverse to the Hamiltonian operator has the following structure
\begin{equation}
  K = \left(
  \begin{array}{cc}
   K_{11} & K_{12} \\
 - K_{12} &  0
  \end{array}
  \right)
\label{K}
\end{equation}
and is defined in (4.5)--(4.8) in \cite{S_Y}. Here (\ref{InvNoeth}) is obtained by applying the operator $K$ to both sides of (\ref{Noether}).

Let us now apply the formula (\ref{InvNoeth}) to determine conserved densities $H^i$ corresponding to all variational symmetries with characteristics $(\varphi^i, \psi^i)$ from the lists given above for the systems $I,II,III$ and $IV$. Using the expression (\ref{K}) for $K$, we rewrite the formula (\ref{InvNoeth}) in an explicit form
\begin{equation}
  \left(\begin{array}{c}
  \delta_u H^i \\ \delta_v H^i  \end{array}
  \right) = \left(\begin{array}{cc}
  K_{11} & K_{12} \\
  -K_{12} & 0
  \end{array}
  \right) \left(\begin{array}{c} \varphi^i
  \\ \psi^i
  \end{array}
  \right)
\label{explicit}
\end{equation}
which provides the formulas determining Hamiltonian densities $H^i$ generating the known symmetries $(\varphi^i, \psi^i)$ from the lists in section \ref{sec-sym}
\begin{equation}
 \delta_u H^i = K_{11}\varphi^i + K_{12}\psi^i,\quad \delta_v H^i = - K_{12} \varphi^i.
\label{sym_H}
\end{equation}

We always start with solving the second equation in (\ref{sym_H}) in which we assume that $H^i$ does not depend on derivatives of $v$, since $\varphi_i$ never contain such derivatives. Hence $\delta_v H^i$ is reduced to the partial derivative $H^i_v$ with respect to $v$, so that the equation
$H^i_v = - K_{12}\varphi_i$ is easily integrated
with respect to $v$ with the "constant of integration" $h^i[u]$ depending only on $u$ and its derivatives. Then the operator $\delta_u$ is applied to the resulting $H^i$, which involves the unknown $\delta_u h^i[u]$, and the result is equated to $\delta_u H^i$ following from the first equation in (\ref{sym_H}) to
determine $\delta_u h^i[u]$. Finally, we reconstruct $h^i[u]$ and hence $H^i$. If we encounter a contradiction, then this particular symmetry is not a variational one and does not lead to an integral.

\subsection{System I}
\label{I.2}
\setcounter{equation}{0}

For the system $I$ formulas \eqref{sym_H} become
\begin{eqnarray}
&& \delta_uH^i = K_{11}\varphi^i + K_{12}\psi^i = \bigl\{v_3D_2 + v_2D_3 + v_{23} + c_4(u_{13}D_2 - u_{23}D_1) \nonumber\\
&&\mbox{} + c_5(u_{22}D_3 - u_{23}D_2) + c_8 (u_{12}D_3 - u_{13}D_2)\bigr\}\varphi^i - u_{23}\psi^i\nonumber\\
&& \delta_vH^i = - K_{12} \varphi^i = u_{23}\varphi^i.
 \label{delHI}
\end{eqnarray}

The solution algorithm for symmetry characteristics in subsection \ref{I.1} of section \ref{sec-sym} yields the following results for Hamiltonian densities
\begin{eqnarray}
&& H^1 = -vu_1u_{23} + \frac{u}{3}\bigl\{c_4(u_{11}u_{23} - u_{12}u_{13}) + c_5(u_{12}u_{23} - u_{22}u_{13})\bigr\}, \nonumber\\
&& H^a = a(z_3)vu_{23} - \frac{a'(z_3)}{2} (c_5u_2^2 + c_8u_1u_2),
 \label{HI}\\
&& H^b = - \frac{b(z_3)}{2}\bigl\{3vu_3u_{23} + c_5(u_2u_3u_{23} - u_3^2u_{22}) + c_8(u_1u_3u_{23} - u_3^2u_{12})\bigr\},\nonumber\\
% \end{eqnarray}
% \begin{eqnarray}
&& H^{(c,e)} = vu_{23}(e(\zeta) - c(\zeta)u_2) + \frac{c_4}{6}(u_1u_{23} - u_2u_{13})(3e(\zeta) - 2c(\zeta)u_2) \nonumber\\
&&\mbox{} - c_8u_1u_{23}(e(\zeta) - c(\zeta)u_2)
% \label{HceI}
\end{eqnarray}
where $\zeta = c_5z_1 - c_8z_2$, whereas $X_2$ and $X_4$ generate non-variational symmetries.

\subsection{System II}
\label{II.2}
\setcounter{equation}{0}

For the system $II$ formulas \eqref{sym_H} become
\begin{eqnarray}
&& \delta_uH^i = \bigl\{a_8(v_2D_1 + D_2v_1) + a_{10}(v_2D_2 + D_2v_2) + a_{11}(v_3D_2 + D_3v_2) \nonumber\\
&&\mbox{} - c_7L_{23(3)} - c_8L_{23(1)}\bigr\}\varphi^i - \Delta\psi^i,\quad \delta_vH^i = \Delta\varphi^i.
 \label{delHII}
\end{eqnarray}

The solution algorithm for the symmetry characteristics in subsection \ref{II.1} ends up with the following results:
$X_1$ is not a variational symmetry,
\begin{eqnarray}
&& H^a = a(z_1)\bigl\{(a_8z_3 + c_8t)\Delta v - \frac{1}{2}(\delta\, u_{23} + c_8a_{10}u_{22})\bigr\}, \nonumber\\
&& H^b = B(v)\Delta
 \label{HII}
\end{eqnarray}
where $B$ is the antiderivative for $b$ ($B'(v)=b(v)$) and we have used the notation \eqref{not}, or explicitly $\Delta=a_8u_{12}+a_{10}u_{22}+a_{11}u_{23}$.
Since determining Hamiltonian density for $\hat X$ is a more complicated problem, we provide here more details of the computation. We use symmetry characteristics \eqref{char6.9} for $\hat\varphi$ and $\hat\psi$ in the formulas \eqref{delHII}. We start with $\delta_v\hat H = \Delta\hat\varphi$ and integrate it with respect to $v$ to obtain
\begin{eqnarray}
&& \hat H = \Delta\bigl\{\hat\Phi(\sigma,\zeta)v -\frac{1}{2}a_8\delta e(\zeta)v^2 -cc_8^2u_1v - \hat G(\sigma,\zeta)u_2v + c_8\delta e(\zeta)u_3v\bigr\} \nonumber\\
&&\mbox{} + \hat h[u].
 \label{hatHII}
\end{eqnarray}
Then we apply to this expression the variational derivative $\delta_u$ and equate the result to the first formula in \eqref{delHII} for $\delta_u\hat H$. All terms containing $v$ are canceled in both sides of this equation and we end up with the following equation for $\hat h[u]$
\begin{eqnarray*}
&& \delta_u\hat h[u] = c_7\left\{(\hat Gu_2-\hat\Phi)_2u_{33} - (\hat Gu_2-\hat\Phi)_3u_{23}\right\} \\
&&\mbox{} + c_8\left\{(\hat Gu_2-\hat\Phi)_2u_{13} - (\hat Gu_2-\hat\Phi)_3u_{12}\right\} \\
&&\mbox{} + c_8^2\left\{[(\delta e + c_7c)u_3]_3u_{12} - [(\delta e + c_7c)u_3]_1u_{23}\right\}
\end{eqnarray*}
with the solution
\begin{eqnarray}
&& \hat h[u] = \frac{\hat G}{2}\left[c_7\left(u_2u_3u_{23} -\frac{1}{2}u_2^2u_{33}\right) + c_8\left(u_2u_3u_{12} -\frac{1}{2}u_2^2u_{13}\right)\right] \nonumber\\
&&\mbox{} + \frac{\hat\Phi}{2}\left[c_7\left(u_2u_{33} - u_3u_{23}\right) + c_8\left(u_2u_{13} - u_3u_{12}\right)\right] \nonumber\\
&&  \mbox{} + c_8^2(\delta e + c_7c)u_1u_3u_{23}.
 \label{hII}
\end{eqnarray}
The sum of the two expressions \eqref{hatHII} and \eqref{hII} presents the Hamiltonian density $\hat H$ which generates the symmetry flow of $\hat X$.
This is the conserved density for the flow of system $II$.

\subsection{System III}
\label{III.2}
\setcounter{equation}{0}

For the system $III$ formulas \eqref{sym_H} become
\begin{eqnarray}
&& \delta_uH^i = \bigl\{2v_3D_3 + v_{33} + c_5(u_{22}D_3 - u_{23}D_2) + c_6(u_{13}D_3 - u_{33}D_1) \nonumber\\
&&\mbox{} + c_7(u_{23}D_3 - u_{33}D_2) + c_8(u_{12}D_3 - u_{13}D_2)\bigr\}\varphi^i - u_{33}\psi^i\nonumber\\
&& \delta_vH^i = u_{33}\varphi^i.
  \label{delHIII}
\end{eqnarray}
It turns out that $X_1$ is not a variational symmetry. To determine the Hamiltonian density $\hat H$ for the symmetry $\hat X$,
we first integrate the second equation in \eqref{delHIII} with $H^i = \hat H$ with respect to $v$ assuming that $\hat H$ depends only on $v$ but not on
derivatives of $v$ with the "constant of integration" $\hat h[u]$ depending only on $u$ and its derivatives
\begin{eqnarray}
&&\hat H = u_{33}\left\langle F(v) - \frac{1}{2} c_8^2b(z_1)v^2 + v\left\{(\delta z_1 + c_8^2z_3)d(\zeta) + g(\zeta) - a(z_1)\right.\right. \nonumber\\
&&\left.\left.\mbox{} - c_6c_8^2b(z_1)u_1 - c_8\bigl(E(\zeta) + c_5c_6b(z_1)\bigr)u_2 \right.\right.\nonumber\\
&&\left.\left.\mbox{} + \bigl[(\delta z_1 + c_8^2z_3)(E'(\zeta) + c_6\delta b(z_1) + \omega(\zeta)\bigr] u_3 \right\}\right\rangle + \hat h[u]
\label{HIII}
\end{eqnarray}
where $F'(v) = f(v)$. To determine $\hat h[u]$, we plug in the variational derivative of $\hat H$ from \eqref{HIII} to the l.h.s. of the first equation \eqref{delHIII}, with an unknown term $\delta_u[\hat h[u]]$, while we utilize $\hat\varphi$ and $\hat\psi$ from \eqref{char6.11} in the r.h.s of the first equation \eqref{delHIII}. We observe that all the terms which depend on $v$ and its derivatives cancel in both sides of the resulting equation, so that
we have only $\delta_u[\hat h[u]]$ remaining on the left and terms depending only on derivatives of $u$ on the right.

The next step is to reconstruct $\hat h[u]$ from its known variational derivative for which we apply the homotopy formula from P. Olver's book \cite{olv}
\begin{equation}
\hat h[u] = \int\limits_0^1 u\,\delta_u\bigl[\hat h[\lambda u]\bigr]\,d\lambda
  \label{homotopy}
\end{equation}
which yields the extra factor $u$ and either $1/3$ or $1/2$ for terms bilinear or linear in $u$, respectively, in the variational derivative
$\delta_u[\hat h[u]]$. We obtain the following result
\begin{eqnarray}
&& \hat h[u] = \frac{u}{3}\bigl\langle c_8E(\zeta)\bigl[c_6(u_{12}u_{33} - u_{13}u_{23}) + c_7(u_{22}u_{33} - u_{23}^2) \nonumber\\
&&\mbox{} - c_8(u_{12}u_{23} - u_{22}u_{13})\bigr] + c_8^2E'(\zeta)\bigl[c_8((u_{12}u_{3} - u_{13}u_{2}) \nonumber\\
&&\mbox{} + c_5(u_{22}u_{3} - u_{23}u_2) + c_7(u_{23}u_3 - u_{33}u_2) \bigr] \nonumber\\
&&\mbox{} + c_6 E'(\zeta)(c_8^2u_{13}u_3 + c_8c_5u_{33}u_2 - \delta u_{33}u_3) \nonumber\\
&&\mbox{} + (\delta z_1 + c_8^2z_3)\left\{E''(\zeta)(c_8^2u_{13} + c_8c_5u_{23} - \delta u_{33})u_3\right.\nonumber\\
&&\left.\mbox{} + E'(\zeta)\bigl[c_8(u_{12}u_{33} - u_{13}u_{23}) + c_5(u_{22}u_{33} - u_{23}^2)\bigr] \right\}\nonumber\\
&&\mbox{} + c_6^2\left\{c_8^2\bigl[b(z_1)(u_{11}u_{33} - u_{13}^2) + b'(z_1)u_1u_{33}\bigr]\right.\nonumber\\
&&\left.\mbox{} + c_8c_5\bigl[2b(z_1)(u_{12}u_{33} - u_{13}u_{23}) + b'(z_1)u_2u_{33}\bigr]\right.\nonumber\\
&&\left.\mbox{} + c_5^2b(z_1)(u_{22}u_{33} - u_{23}^2) - \delta b'(z_1)u_3u_{33}\right\}\nonumber\\
&&\mbox{} + \omega(\zeta)\bigl[c_8(u_{12}u_{33} - u_{13}u_{23}) + c_5(u_{22}u_{33} - u_{23}^2)\bigr]\nonumber\\
&&\mbox{} + \omega'(\zeta)(c_8^2u_{13} + c_8c_5u_{23} - \delta u_{33})u_3\bigr\rangle
  \label{hat hIII}\\
&&\mbox{} + \frac{u}{2}\left\{d(\zeta)\bigl[c_8^2(c_8u_{12} + c_5u_{22} + c_7u_{23} + c_6u_{13}) - c_6\delta u_{33}\bigr]\right.\nonumber\\
&&\left.\mbox{} + \left[(\delta z_1 + c_8^2z_3)d^{\,\prime}(\zeta) + g'(\zeta)\right](c_8^2u_{13} + c_8c_5u_{23} - \delta u_{33}) + c_6a'(z_1)u_{33}\right\} \nonumber
\end{eqnarray}
where primes denote derivatives.

The sum of expressions \eqref{HIII} and \eqref{hat hIII} yields the Hamiltonian density for the symmetry $\hat X$ which is conserved by the flow of the system $III$.

\subsection{System IV}
\label{IV.2}
\setcounter{equation}{0}

For the system $IV$ formulas \eqref{sym_H} become
\begin{eqnarray}
&& \delta_uH^i = \bigl\{a_7(2v_1D_1 + v_{11}) + a_8(v_2D_1 + v_1D_2 + v_{12}) +  \nonumber\\
&&\mbox{} + a_9(v_3D_1 + v_1D_3 + v_{13}) + c_1(u_{11}D_2 - u_{12}D_1) \nonumber\\
&&\mbox{} + c_3(u_{12}D_2 - u_{22}D_1) + c_4(u_{13}D_2 - u_{23}D_1)\bigr\}\varphi^i \nonumber\\
&&\mbox{} - (a_7u_{11} + a_8u_{12} + a_9u_{13})\psi^i\nonumber\\
&&\delta_vH^i = (a_7u_{11} + a_8u_{12} + a_9u_{13})\varphi^i.
 \label{delHIV}
\end{eqnarray}
We apply the solution algorithm at the beginning of this section to the symmetry characteristics in subsection \ref{IV.1} of section \ref{sec-sym}. Introduce the shorthand notation $\Delta = a_7u_{11} + a_8u_{12} + a_9u_{13}$.

Symmetry $X_1$ is not a variational symmetry.

For the symmetry $Y_a$, we start with the relation \[\delta_vH^a = \Delta\varphi_a = \Delta a(z_3,v).\]
Introducing $A(z_3,v)$ as the antiderivative of $a$, $a(z_3,v) = A_v(z_3,v)$, we integrate the last equation with respect to $v$ to obtain
\begin{equation}
 H^a = (a_7u_{11} + a_8u_{12} + a_9u_{13})A(z_3,v) + h^a[u].
 \label{Ha}
\end{equation}
Next, we calculate the variational derivative $\delta_uH^a$ containing yet unknown term $\delta_uh^a[u]$ and equate it to the expression for $\delta_uH^a$
from \eqref{delHIV} with $\varphi^i = \varphi_a$ and $\psi^i = \psi_a$. The resulting equation can be satisfied only if $a=a(v)$ is independent of $z_3$, same as $A=A(v)$. Then it follows that $\delta_uh^a[u] = 0$, so that we can choose $h^a[u] = 0$ and from \eqref{Ha} we have
\begin{equation}
 H^a = (a_7u_{11} + a_8u_{12} + a_9u_{13})A(v),\qquad A'(v) = a(v).
 \label{H_a}
\end{equation}

For the symmetry $\hat X$, we start again with the second equation in \eqref{delHIV}
\begin{eqnarray*}
 && \delta_v\hat H  = \Delta\hat\varphi  = \Delta \bigl[F(\sigma) - \Omega(\zeta) - a_9\alpha E(\zeta)v - \bigl(G(\sigma) - c_4\beta E(\zeta)\bigr)u_1\\
&&\mbox{} + c_4\alpha E(\zeta)u_2\bigr].
\end{eqnarray*}
Integrating this equation in $v$ we obtain
\begin{eqnarray}
&& \hat H = (a_7u_{11} + a_8u_{12} + a_9u_{13})\left\{-\frac{a_9}{2} \alpha E(\zeta)v^2 + v\Bigl[F(\sigma) - \Omega(\zeta)\right. \nonumber\\
&&\left.\mbox{} - \bigl(G(\sigma) - c_4\beta E(\zeta)\bigr)u_1 \Bigr] + c_4\alpha E(\zeta)u_2\right\} + \hat h[u]
 \label{hatH}
\end{eqnarray}
with $\hat h[u]$ as a ``constant'' of integration. We compute the variational derivative $\delta_u\hat H$ of \eqref{hatH} and equate it to the expression for $\delta_u\hat H$ from \eqref{delHIV} with $\varphi^i = \hat\varphi$ and $\psi^i = \hat\psi$ taken from \eqref{char6.13}. Then all the terms containing $v$ and its derivatives are canceled on both sides of the resulting equation and we end up with the equation
\begin{eqnarray}
&& \delta_u\hat h[u] = (c_1u_{12} + c_3u_{22} + c_4u_{23})\Bigl[\alpha\bigl(G'(\sigma)u_1 - F'(\sigma)\bigr)\nonumber\\
&&\mbox{} + \bigl(G(\sigma) - c_4\beta E(\zeta)\bigr)u_{11} - c_4\alpha E(\zeta)u_{12}\Bigr]\nonumber\\
&&\mbox{} + (c_1u_{11} + c_3u_{12} + c_4u_{13})\Bigl[\beta\bigl(G'(\sigma)u_1 - F'(\sigma)\bigr)\nonumber\\
&&\mbox{} + c_4\bigl(c_4\beta E'(\zeta)u_1 - \Omega'(\zeta)\bigr) - \bigl(G(\sigma) - c_4\beta E(\zeta)\bigr)u_{12}\nonumber\\
&&\mbox{} + c_4\alpha\bigl(c_4E'(\zeta)u_2 + E(\zeta)u_{22}\bigr)\Bigr]
 \label{delt hat h}
\end{eqnarray}
We reconstruct $\hat h[u]$ from \eqref{delt hat h} using the homotopy formula \eqref{homotopy}.
It modifies \eqref{delt hat h} by the extra factor $u$ and either $1/3$ or $1/2$ for the terms that bilinear or linear in $u$, respectively, in the variational derivative $\delta_u[\hat h[u]]$. Thus, we obtain the following result
\begin{eqnarray}
&& \hat h[u] = u(c_1u_{12} + c_3u_{22} + c_4u_{23})\left\{\frac{1}{3}\Bigl[\alpha G'(\sigma)u_1 \right.\nonumber\\
&&\left.\mbox{} + \bigl(G(\sigma) - c_4\beta E(\zeta)\bigr)u_{11} - c_4\alpha E(\zeta)u_{12}\Bigr] - \frac{1}{2}\alpha F'(\sigma)\right\} \nonumber\\
&&\mbox{} + u(c_1u_{11} + c_3u_{12} + c_4u_{13})\left\{\frac{1}{3}\Bigl[\bigl(\beta G'(\sigma) + c_4^2\beta E'(\zeta)\bigr)u_1\right. \nonumber\\
&&\left.\mbox{} - \bigl(G(\sigma) - c_4\beta E(\zeta)\bigr)u_{12} + c_4\alpha\bigl(c_4E'(\zeta)u_2 + E(\zeta)u_{22}\bigr)\Bigr]\right. \nonumber\\
&&\left.\mbox{} - \frac{1}{2}\bigl(\beta F'(\sigma) + c_4\Omega'(\zeta)\bigr)\right\}.
 \label{hat h}
\end{eqnarray}
Using this result in \eqref{hatH}, we obtain the Hamiltonian density generating the symmetry flow of $\hat X$ which is conserved by the system $IV$.

Since we have eliminated from $\hat X$ in \eqref{hatsymIV} the obvious translational symmetries $X_2=-\partial_2$ and $X_3=-\partial_3$, for completeness we present below the Hamiltonian densities $H^2$ and $H^3$ for these symmetries, which are conserved by the system $IV$
\begin{eqnarray}
&& H^2 = vu_2(a_7u_{11} + a_8u_{12} + a_9u_{13}) + \frac{1}{3}\bigl[c_1u_2(u_1u_{12} - u_2u_{11})\nonumber\\
&&\mbox{} - c_4u_1(u_3u_{22} - u_2u_{23}) \bigr],\nonumber\\
&& H^3 = vu_3(a_7u_{11} + a_8u_{12} + a_9u_{13}) + \frac{1}{3}\bigl[c_1u_3(u_1u_{12} - u_2u_{11})\nonumber\\
&&\mbox{} + c_3u_1(u_3u_{22} - u_2u_{23}) \bigr].
 \label{H23}
\end{eqnarray}

Concerning our results for conservation laws for systems $III$ and $IV$ where we have used the homotopy formula, we should note that the conserved densities are by no means unique, so that we could add or subtract total divergences to them in order to obtain more compact expressions.

\section{Recursion operators and hierarchies of the new bi-Hamiltonian systems}
\setcounter{equation}{0}
\label{sec-hier}

So far, we have not used the Magri integrability \cite{magri} of new bi-Hamiltonian systems studied above.
Now we will consider hierarchies of our systems $I,II,III$ and $IV$  related to their bi-Hamiltonian property.
We first review main properties of hierarchies of bi-Hamiltonian systems (see, e.g., \cite{ff} and \cite{sheftel}) with some new features related to our use of inverse recursion operators.

Any bi-Hamiltonian system has the form
\begin{equation}
\left(
\begin{array}{c}
 u_t\\
 v_t
\end{array}
\right) = J_0
\left(
\begin{array}{c}
 \delta_u H_1\\
 \delta_v H_1
\end{array}
\right) = J_1
\left(
\begin{array}{c}
 \delta_u H_0\\
 \delta_v H_0
\end{array}
\right)
  \label{be_Ham}
\end{equation}
where $J_0$ and $J_1$ are compatible Hamiltonian operators which determine the structures of Poisson brackets, $H_1$ and $H_0$ being the corresponding Hamiltonian densities, respectively, $\delta_u, \delta_v$ are Euler-Lagrange operators \cite{olv}.

Hamiltonian operators are skew-symmetric: $J^\dagger = - J$, where $\dagger$ denotes the (formal) adjoint operator, and they satisfy the Jacobi identities. The compatibility of $J_0$ and $J_1$ requires the Jacobi identities to hold also for linear combinations of these operators with arbitrary constant coefficients.
A check of the Jacobi identities and compatibility of the two Hamiltonian structures $J_0$ and $J_1$ is straightforward but too lengthy to be presented here.
Therefore, we restrict ourselves here by demonstrating that all our candidates for Hamiltonian operators are indeed manifestly skew-symmetric.
The method of the functional multi-vectors for checking the Jacobi identity and the compatibility of the Hamiltonian operators is developed by P. Olver
in \cite{olv}, chapter 7, that we have recently applied for checking bi-Hamiltonian structure of the general heavenly equation \cite{sym} and the
first heavenly equation of Pleba\'nski \cite{sy_tri} under the well-founded conjecture that this method is applicable for nonlocal Hamiltonian operators as well.
We note that our operators $L_{ij(k)}$ are also skew-symmetric.

A recursion operator $R$ maps any symmetry again into a symmetry. Operator $R$ provides a relation $J_1=RJ_0$ between Hamiltonian operators in the bi-Hamiltonian representation \eqref{be_Ham}, so that $R$ admits the symplectic-implectic factorization
$R=J_1J_0^{-1}$. Here $K=J_0^{-1}$ is a symplectic operator and ``implectic'' is another name for Hamiltonian operator. Fuchssteiner and Fokas \cite{ff} showed that if a recursion operator has the form $R=J_1J_0^{-1}$, where $J_0$ and $J_1$ are compatible Hamiltonian operators, then it is hereditary (Nijenhuis).
In order that the repeated applications of the adjoint of a hereditary recursion operator to a vector of variational derivatives of an integral produce again vectors of variational derivatives of (another) integral, it is necessary (but not sufficient) that the result of the \textit{first} such application will be a vector of variational derivatives (see e.g. Hilfssatz 4 c) in \cite{Oe}). This condition is satisfied in our case because the equation \eqref{be_Ham} can be rewritten in the form
\begin{equation}
\left(
\begin{array}{c}
 \delta_u H_1\\
 \delta_v H_1
\end{array}
\right) = R^\dagger
\left(
\begin{array}{c}
 \delta_u H_0\\
 \delta_v H_0
\end{array}
\right)
  \label{R^+}
\end{equation}
where $R^\dagger = J_0^{-1}J_1$. Therefore, applying $R^\dagger$ we can determine the next Hamiltonian density $H_2$ in the hierarchy from the equation
\begin{equation}
\left(
\begin{array}{c}
 \delta_u H_2\\
 \delta_v H_2
\end{array}
\right) = R^\dagger
\left(
\begin{array}{c}
 \delta_u H_1\\
 \delta_v H_1
\end{array}
\right) = (R^\dagger)^2
\left(
\begin{array}{c}
 \delta_u H_0\\
 \delta_v H_0
\end{array}
\right)
  \label{H2}
\end{equation}
and so on. More generally, we have
\begin{equation}
\left(
\begin{array}{c}
 \delta_u H_n\\
 \delta_v H_n
\end{array}
\right) = (R^\dagger)^n
\left(
\begin{array}{c}
 \delta_u H_0\\
 \delta_v H_0
\end{array}
\right).
  \label{Hn}
\end{equation}
Taking the adjoint of $J_1=RJ_0$ we have
\begin{equation}
 J_1 = J_0R^\dagger\quad \Rightarrow \quad J_n = J_0({R^\dagger})^n
 \label{J_n}
\end{equation}
and as a consequence of \eqref{Hn} and \eqref{J_n}
\begin{equation}
\left(\begin{array}{c}
 u_{\tau_{m+n}} \\
 v_{\tau_{m+n}}
\end{array}\right) =
J_{m}
\left(
\begin{array}{c}
 \delta_u H_n\\
 \delta_v H_n
\end{array}
\right) = J_{n}
\left(
\begin{array}{c}
 \delta_u H_m\\
 \delta_v H_m
\end{array}
\right) = J_{k}\left(
\begin{array}{c}
 \delta_u H_l\\
 \delta_v H_l
\end{array}
\right)
  \label{Hmn}
\end{equation}
where $m+n = k+l$ and $m,n,k,l$ are nonnegative integers.
We will see that sometimes, in order to generate nonlocal (higher) flows in an hierarchy and even to obtain bi-Hamiltonian representation, we also need the inverse recursion operator $R^{-1}$ which satisfies the relations $RR^{-1}=R^{-1}R=I$ where $I$ is the unit operator.  If we have
\begin{equation}
R=\left(\begin{array}{cc}
   a & b\\
   c & d
\end{array}
\right)
\label{R}
\end{equation}
with noncommuting entries, then the inverse operator is determined by the formula
\begin{equation}
   R^{-1} = \left(\begin{array}{cc}
   e & f\\
   g & h
\end{array}
\right) = \left(\begin{array}{cc}
   (a-bd^{-1}c)^{-1}, & (c-db^{-1}a)^{-1}\\
   (b-ac^{-1}d)^{-1}, & (d-ca^{-1}b)^{-1}
\end{array}
\right)
 \label{R^-1}
\end{equation}
which we derived earlier in \cite{sym} in a slightly different context. Here each operator $x^{-1}$ can make sense merely as a \textit{formal} inverse of $x$.

A proper way to deal with inversion of operators in total derivatives like $W$ (below) is through the theory of differential coverings, see e.g. the reference \cite{Kras} and references therein.
Specifically for the inversion of recursion operators and the proper definition of their action, see also \cite{Gu,Ma}.

We can always properly define the inverse operators in a similar way as we did in \cite{sym}, so that $xx^{-1}=x^{-1}x=I$.

Using $R^{-1}$, we define $J_{-1} = R^{-1}J_0 = J_0(R^{-1})^\dagger$, so that $J_0=RJ_{-1}$ and $J_0=R^{-1}J_1=J_1(R^{-1})^\dagger$. By virtue of \eqref{R^+}
\begin{equation}
\left(
\begin{array}{c}
 \delta_u H_0\\
 \delta_v H_0
\end{array}
\right) = (R^{-1})^\dagger
\left(
\begin{array}{c}
 \delta_u H_1\\
 \delta_v H_1
\end{array}
\right)
  \label{R^{-1+}}
\end{equation}
we have bi-Hamiltonian representation in the form
\begin{equation}
J_{-1}
\left(
\begin{array}{c}
 \delta_u H_1\\
 \delta_v H_1
\end{array}
\right) = J_0
\left(
\begin{array}{c}
 \delta_u H_0\\
 \delta_v H_0
\end{array}
\right).
 \label{J-1}
\end{equation}
More generally, we define
$J_{-m}= R^{-m}J_0 = J_0(R^{-m})^\dagger$ for positive integer $m$ which implies
\begin{equation}
J_{-m}
\left(
\begin{array}{c}
 \delta_u H_n\\
 \delta_v H_n
\end{array}
\right) = J_{n-m}
\left(
\begin{array}{c}
 \delta_u H_0\\
 \delta_v H_0
\end{array}
\right) = J_{n-k-m}
\left(
\begin{array}{c}
 \delta_u H_k\\
 \delta_v H_k
\end{array}
\right),
  \label{J-m}
\end{equation}
e.g., for $n=4$, $m=1$, $k=2$ this yields
\begin{equation*}
J_{-1}
\left(
\begin{array}{c}
 \delta_u H_4\\
 \delta_v H_4
\end{array}
\right) = J_1
\left(
\begin{array}{c}
 \delta_u H_2\\
 \delta_v H_2
\end{array}
\right).
\end{equation*}
Relations \eqref{Hmn} are now valid for any integer $m,n,k,l$ satisfying  $m+n = k+l$, including their negative values.

There is a fairly extensive literature on negative ("minus first") flows, e.g. \cite{Anco} and references therein.

\subsection{Hierarchy of system $I$}
\label{I.3}
\setcounter{equation}{0}

Recursion operator for the system $I$, as given in \cite{S_Y}, with $a_{11}=1$ and the notation $W = c_8L_{13(2)} + c_5L_{23(2)}$ has the form
\begin{equation}
 R = \left(
\begin{array}{cc}
  - L_{12(3)}^{-1}(W - v_2D_3),                              & - L_{12(3)}^{-1}u_{23} \\
 \begin{array}{c}
  \displaystyle\frac{1}{c_8u_{23}} \bigl[c_8(c_8-c_4)v_2D_3 + c_9W\bigr] \\ \hspace*{5mm}
 - \displaystyle\frac{v_3}{u_{23}} D_2 L_{12(3)}^{-1} (W - v_2D_3),
\end{array}        &\displaystyle -\frac{v_3}{u_{23}}D_2 L_{12(3)}^{-1}u_{23} + c_4 - c_8
\end{array}\right)
  \label{RI}
\end{equation}
where the relation $c_8c_{10} = c_5c_9$ has been used. The first Hamiltonian operator for system $I$ reads \cite{S_Y}
\begin{equation}
 J_0 = \frac{1}{u_{23}} \left(
 \begin{array}{cc}
 0  & 1 \\
 -1 & K_{11}\displaystyle\frac{1}{u_{23}}
\end{array}\right)
  \label{J0I}
\end{equation}
where $K_{11} = v_3D_2 + D_3v_2 - \bigl[(c_4-c_8)L_{12(3)} + W\bigr]$.
The corresponding Hamiltonian density reads
\begin{equation}
 H_1 = \frac{v^2}{2} u_{23} + \frac{c_9}{3c_8} uW[u_1]
 \label{H_1I}
\end{equation}
with the variational derivatives
\[\delta_uH_1 = D_2(vv_3) + \frac{c_9}{c_8} W[u_1],\quad \delta_vH_1 = vu_{23}.\]
In the bi-Hamiltonian representation \eqref{be_Ham} for the system $I$ we have
the second Hamiltonian operator
\begin{equation}
 J_1 = \left(\begin{array}{cc}
  L_{12(3)}^{-1}             & - \left(L_{12(3)}^{-1}D_2v_3 + c_8-c_4\right)\displaystyle\frac{1}{u_{23}} \\
  \displaystyle\frac{1}{u_{23}}\left(v_3D_2L_{12(3)}^{-1} + c_8-c_4\right) & J_1^{22}
 \end{array}\right)
 \label{J_1}
\end{equation}
where the entry $J_1^{22}$ is defined by
\begin{eqnarray}
 && J_1^{22} = \frac{1}{u_{23}}(c_9L_{13(2)} + c_{10}L_{23(2)})\frac{1}{u_{23}} - \frac{v_3}{u_{23}}D_2L_{12(3)}^{-1}D_2\frac{v_3}{u_{23}}
 \label{J_1^22}
\\
&&\mbox{} + \frac{c_4-c_8}{u_{23}}\left\{D_2v_3 + v_3D_2
 - (c_4L_{12(3)} + c_5L_{23(2)} + c_8L_{23(1)})\right\}\frac{1}{u_{23}}.\nonumber
\end{eqnarray}
Here $J_1$ is manifestly skew-symmetric. The corresponding Hamiltonian density is determined by the formula \eqref{R^{-1+}}
\begin{equation}
 H_0 = - k\left\{\frac{v^2}{2} + \frac{c_9}{2c_8}\left[2u_1v + (c_4-c_8)u_1^2\right]\right\}u_{23}
 \label{H0I}
\end{equation}
where $k = \displaystyle\frac{c_8}{[c_8(c_8-c_4) + c_9]}$\,.
To obtain the next Hamiltonian density in the hierarchy of system $I$, we use the relation \eqref{H2}
\begin{equation}
\left(
\begin{array}{c}
 \delta_u H_2\\
 \delta_v H_2
\end{array}
\right) = R^\dagger
\left(
\begin{array}{c}
 \delta_u H_1\\
 \delta_v H_1
\end{array}
\right)
  \label{delH2I}
\end{equation}
with the result
\begin{eqnarray}
&& H_2 = \left\{\frac{1}{2}(c_4-c_8)v^2 + (c_9u_1 + c_{10}u_2)v\right\}u_{23}
  \label{H2I}\\
&&\mbox{} - \frac{c_8}{3}\left\{c_9u_2(u_1u_{13} - u_3u_{11}) + c_{10}u_1(u_3u_{22} - u_2u_{23})\right\}. \nonumber
\end{eqnarray}
The corresponding Hamiltonian flow
\begin{equation}
\left(
\begin{array}{c}
 u_{\tau_2}\\
 v_{\tau_2}
\end{array}
\right) = J_0
\left(
\begin{array}{c}
 \delta_u H_2\\
 \delta_v H_2
\end{array}
\right) = J_1
\left(
\begin{array}{c}
 \delta_u H_1\\
 \delta_v H_1
\end{array}
\right)
  \label{H2Iflow}
\end{equation}
is the local one
\begin{eqnarray}
&& u_{\tau_2} = (c_4-c_8)v + c_9u_1 + c_{10}u_2\nonumber\\
&& v_{\tau_2} = (c_4-c_8)q + c_9v_1 + c_{10}v_2
 \label{t2Ifl}
\end{eqnarray}
where $q$ is defined in \eqref{6.5}. This flow is generated by the following combination of symmetry generators \eqref{sym6.5}, \eqref{XinfI }
$X^{(1)} = - c_9X_1 - c_{10}X_{(c=1,e=0)} - (c_4-c_8)X_3$.

Another flow is generated by $H_2$ via the next Hamiltonian operator $J_2=J_1R^\dagger = J_0(R^\dagger)^2$ in the form
\begin{equation}
\left(
\begin{array}{c}
 u_{\tau_3}\\
 v_{\tau_3}
\end{array}
\right) = J_1
\left(
\begin{array}{c}
 \delta_u H_2\\
 \delta_v H_2
\end{array}
\right) = J_2
\left(
\begin{array}{c}
 \delta_u H_1\\
 \delta_v H_1
\end{array}
\right)
  \label{H2J2_Iflow}
\end{equation}
where we can avoid the explicit use of operator $J_2$. The explicit form of the flow
\begin{eqnarray}
&& u_{\tau_3} = \{(c_4-c_8)^2 - c_9\}v + c_9(c_4-2c_8)u_1 + (c_4c_{10} - 2c_5c_9)u_2\nonumber\\
&& v_{\tau_3} = \{(c_4-c_8)^2 - c_9\}q + c_9(c_4-2c_8)v_1 + (c_4c_{10} - 2c_5c_9)v_2\nonumber\\
 \label{t2IJ1H2}
\end{eqnarray}
show that it is still a local one. This flow is generated by the following combination of symmetry generators \eqref{sym6.5}, \eqref{XinfI }
$X^{(2)} = (c_4 - c_8)X^{(1)} + c_9(X_3 + c_5 X_{(c=-1,e=0)})$.

Since we are looking for nonlocal (higher) flows, we continue applying powers of the adjoint recursion operator
\begin{equation}
\left(
\begin{array}{c}
 \delta_u H_3\\
 \delta_v H_3
\end{array}
\right) = R^\dagger
\left(
\begin{array}{c}
 \delta_u H_2\\
 \delta_v H_2
\end{array}
\right)
  \label{delH3I}
\end{equation}
which yields the next Hamiltonian density in the hierarchy
\begin{eqnarray}
&& H_3 = \left\{\frac{1}{2}\left[(c_4-c_8)^2 - c_9\right]v^2 + (c_4-2c_8)(c_9u_1 + c_{10}u_2)v\right\}u_{23}\nonumber\\
&&\mbox{} + \frac{1}{3}(c_8^2 - c_9)\left\{c_9u_2(u_1u_{13} - u_3u_{11}) + c_{10}u_1(u_3u_{22} - u_2u_{23})\right\}.
  \label{H3I}
\end{eqnarray}
The corresponding Hamiltonian flow
\begin{equation}
\left(
\begin{array}{c}
 u_{\tau_4}\\
 v_{\tau_4}
\end{array}
\right) = J_1
\left(
\begin{array}{c}
 \delta_u H_3\\
 \delta_v H_3
\end{array}
\right) = J_2
\left(
\begin{array}{c}
 \delta_u H_2\\
 \delta_v H_2
\end{array}
\right)
  \label{H3J1_Iflow}
\end{equation}
has the explicit form
\begin{eqnarray}
&&\hspace*{-28pt} u_{\tau_4} = \{(c_4-c_8)^3 + c_9(3c_8-2c_4)\}v + \{(c_4^2 - c_9 + 3c_8(c_8-c_4)\}(c_9u_1 + c_{10}u_2)\nonumber\\
&&\hspace*{-28pt} v_{\tau_4} = \{(c_4-c_8)^3 + c_9(3c_8-2c_4)\}q + \{(c_4^2 - c_9 + 3c_8(c_8-c_4)\}(c_9v_1 + c_{10}v_2)\nonumber\\
 \label{t4J1H3}
\end{eqnarray}
which is again local.

Thus, applying positive powers of $R^\dagger$ we obtain only local flows of point symmetries with a similar dependence on $u$ and $v$ and transformed coefficients.
Hence, to obtain nonlocal (or higher) flows we need the \textit{inverse recursion operator} $R^{-1}$ which will allow us to move along the hierarchy in the opposite direction.
Let $R=\left(\begin{array}{cc}
a & b\\
c & d \end{array}\right)$ and
$R^{-1} = \left(\begin{array}{cc}
e & f\\
g & h \end{array}\right)$.
The solution to equations $RR^{-1} = R^{-1}R = I$ is given by the formula \eqref{R^-1} where the values of the entries $a,b,c,d$ are given in the formula \eqref{RI} for $R$. However, we will use here more simple formulas
\begin{equation}
f=(c-db^{-1}a)^{-1},\quad e=-fdb^{-1},\quad h=-b^{-1}af,\quad g=-hca^{-1}
 \label{R^-1simple}
\end{equation}
equivalent to \eqref{R^-1}. We obtain the result
\begin{eqnarray}
&&\hspace*{-9pt} R^{-1} = k\left\{\left(
\begin{array}{cc}
 - W^{-1}\bigl[(c_8-c_4)L_{12(3)}+v_3D_2\bigr], & W^{-1}u_{23} \\
 - \displaystyle\frac{v_2}{u_{23}}D_3W^{-1}\bigl[(c_8-c_4)L_{12(3)}+v_3D_2\bigr], & \displaystyle\frac{v_2}{u_{23}}D_3W^{-1}u_{23}
\end{array}
\right)\right.\nonumber\\
&&\hspace*{-9pt}\left.\mbox{} + \left(\begin{array}{cc}
         0 ,               &               0 \\
   \displaystyle\frac{1}{c_8u_{23}}(c_8v_3D_2 - c_9L_{12(3)}), & -1
\end{array}\right)\right\}
 \label{RIent}
\end{eqnarray}
where $k = \displaystyle\frac{c_8}{[c_8(c_8-c_4) + c_9]}$\,. Using its adjoint $(R^{-1})^\dagger$, we define the Hamiltonian operator $J_{-1}=J_0(R^{-1})^\dagger$ in the form
\begin{equation}
 J_{-1} = k\left(\begin{array}{cc}-W^{-1},  &\displaystyle (W^{-1}D_3v_2 - 1)\frac{1}{u_{23}} \\
\displaystyle\frac{1}{u_{23}}(1 - v_2D_3W^{-1}), & J_{-1}^{22}
\end{array}\right)
\label{J-1I}
\end{equation}
where
\begin{equation}
J_{-1}^{22} = \frac{1}{u_{23}}\left[v_2D_3W^{-1}D_3v_2 - (D_3v_2+v_2D_3) + W - \frac{1}{k}L_{12(3)}\right]\frac{1}{u_{23}}.
 \label{J-1^22}
\end{equation}
Here $J_{-1}$ is manifestly skew-symmetric.
Now we can consider the Hamiltonian flow
\begin{equation}
\left(
\begin{array}{c}
 u_{\tau_{-1}}\\
 v_{\tau_{-1}}
\end{array}
\right) = J_{-1}
\left(
\begin{array}{c}
 \delta_u H_0\\
 \delta_v H_0
\end{array}
\right)
\label{t_-1}
\end{equation}
with $H_0$ defined in \eqref{H0I}, or in an explicit form
\begin{eqnarray}
&& u_{\tau_{-1}} = k^2\left\{- \frac{c_9}{c_8} W^{-1}L_{13(2)}\bigl[v + (c_4-c_8)u_1\bigr] + v + \frac{c_9}{c_8} u_1)\right\}\nonumber\\
&& v_{\tau_{-1}} = \frac{k^2}{u_{23}} \left\{- \frac{c_9}{c_8}v_2D_3W^{-1}L_{13(2)}\bigl[v + (c_4-c_8)u_1\bigr]\right.\nonumber\\
&&\left.\mbox{} + \frac{c_9}{c_8}L_{13(2)}\bigl[v + (c_4-c_8)u_1\bigr] + \frac{1}{k}L_{12(3)}\left[v + \frac{c_9}{c_8}u_1\right]\right.
\label{fl_1I}\\
&&\left.\mbox{} - W\left[v + \frac{c_9}{c_8}u_1\right] + v_2\left(v_3 + \frac{c_9}{c_8}u_{13}\right)\right\}. \nonumber
\end{eqnarray}
Due to $W^{-1}$, this is a nonlocal flow, so that its stationary solutions $u_{\tau_{-1}} = v_{\tau_{-1}} = 0$ need not to admit the reduction in the number of independent variables.

The equations \eqref{fl_1I} imply
\begin{eqnarray}
&& v_{\tau_{-1}} = \frac{v_2}{u_{23}}D_3u_{\tau_{-1}} + \frac{k^2}{u_{23}}\left\{\frac{c_9}{c_8}L_{13(2)}[v + (c_4 - c_8)u_1]\right.\nonumber\\
&&\left.\mbox{} - W\left[v + \frac{c_9}{c_8}u_1\right] + \frac{1}{k}L_{12(3)}\left[v + \frac{c_9}{c_8}u_1\right]\right\}.
 \label{compact}
\end{eqnarray}
 For stationary solutions from \eqref{fl_1I} we have
\begin{equation}
 W[c_8v+c_9u_1] = c_9L_{13(2)}[v + (c_4-c_8)u_1]
\label{u tau=0}
\end{equation}
and \eqref{compact} implies
\begin{equation}
  L_{12(3)}[c_8v+c_9u_1] = 0.
 \label{v_tau=0}
\end{equation}

Further analysis is needed to determine an explicit solution to these equations which we postpone for future publications.

Due to the greater simplicity of a similar problem for the system $II$, we defer to the next subsection the detailed exposition of the procedure of studying the compatibility of the original bi-Hamiltonian system and its first nonlocal flow, which shows that both flows commute and hence the latter flow is indeed a nonlocal symmetry of the first flow. We will also show there how to treat the corresponding stationary equations.

\subsection{Hierarchy of system $II$}
\label{II.3}
\setcounter{equation}{0}

The system $II$ has the form \eqref{6.9} where we have used the notation \eqref{not}. Recursion operator has the form \cite{S_Y}
\begin{eqnarray}
R = \left(\begin{array}{cc}
           -L_{23(t)}^{-1}v_2\hat{\Delta}, & L_{23(t)}^{-1}\Delta \\
   \displaystyle -\frac{q}{v_2}D_2L_{23(t)}^{-1}v_2\hat{\Delta}+\hat{c}, &\displaystyle \frac{1}{v_2}\left\{qD_2L_{23(t)}^{-1}\Delta - \hat{c}[u_2]\right\}
\end{array}\right).
\label{RII}
\end{eqnarray}
The first Hamiltonian operator reads
\begin{equation}
 J_0 =\left(\begin{array}{cc}
  0,  & \Delta^{-1}\\
  - \Delta^{-1}, & \Delta^{-1}K_{11}\Delta^{-1}
 \end{array}
 \right)
\label{J_02}
\end{equation}
where $K_{11}=v_2\hat{\Delta}+D_2\hat{\Delta}[v] - \hat c[u_3]D_2 + \hat c[u_2]D_3$.
The second Hamiltonian operator reads
\begin{equation}
 J_1 = RJ_0 =\left(\begin{array}{cc}
  -L_{23(t)}^{-1},  &\displaystyle (L_{23(t)}^{-1}D_2q\Delta -\hat{c}[u_2])\frac{1}{v_2\Delta} \\[2pt]
  - \displaystyle\frac{1}{v_2\Delta}(q\Delta D_2L_{23(t)}^{-1}-\hat{c}[u_2]), & J_1^{22}
 \end{array}
 \right)
\label{J_12}
\end{equation}
where
\begin{eqnarray}
&& J_1^{22} = \hat{c}\frac{1}{\Delta} - \hat{c}[u_2]\frac{1}{\Delta}\hat{\Delta}\frac{1}{\Delta} + \frac{q}{v_2}D_2L_{23(t)}^{-1}D_2\frac{q}{v_2}
- \frac{q}{v_2}D_2\frac{\hat{c}[u_2]}{v_2\Delta}\nonumber\\
&&\mbox{} -  \frac{\hat{c}[u_2]}{v_2\Delta}D_2\frac{q}{v_2} + \frac{\hat{c}[u_2]}{v_2\Delta}L_{23(t)}\frac{\hat{c}[u_2]}{v_2\Delta}
\label{J1_22}
\end{eqnarray}
and $q=v_t$ is given by the r.h.s. \eqref{6.9} of system $II$. Formulas \eqref{J_12} and \eqref{J1_22} show that $J_1$ is manifestly skew-symmetric:
$J_1^\dagger = - J_1$.
The Hamiltonian density corresponding to $J_0$ reads
\begin{equation}
H_1 = \frac{v^2}{2}\Delta = \frac{v^2}{2} (a_8u_{12} + a_{10}u_{22} + a_{11}u_{23}).
\label{H1_2}
\end{equation}
However, there is a problem with the Hamiltonian density $H_0$ corresponding to $J_1$, related to the fact that $v$ belongs to the kernel of the operator $L_{23(t)}$, so that to enforce the relation $L_{23(t)}^{-1}L_{23(t)} = 1$ we had to skip $v$ which is needed to reproduce the correct second equation in \eqref{be_Ham}.

To determine the correct $H_0$ we apply the relation
\begin{equation}
 \left(\begin{array}{c}
 \delta_u H_0 \\
 \delta_v H_0
\end{array}\right) = (R^\dagger)^{-1}
\left(\begin{array}{c}
 \delta_u H_1 \\
 \delta_v H_1
\end{array}\right)
 \label{R^+^-1II}
\end{equation}
using an adjoint inverse recursion operator, inverse to $R^\dagger$.
Operator $R^\dagger$ reads
\begin{eqnarray}
R^\dagger = \left(\begin{array}{cc}
           - \hat{\Delta}v_2L_{23(t)}^{-1}, &\displaystyle \hat{\Delta}v_2L_{23(t)}^{-1}D_2\frac{q}{v_2} - \hat{c} \\
    - \Delta L_{23(t)}^{-1} , &\displaystyle \left\{\Delta L_{23(t)}^{-1}D_2q - \hat{c}[u_2]\right\}\frac{1}{v_2}
\end{array}\right).
\label{R^+II}
\end{eqnarray}
The inverse operator is determined by the formula
\begin{eqnarray}
\hspace*{-2pt}&& R^{-1} = \nonumber\\
\hspace*{-2pt}&& \left(\begin{array}{cc}
           W^{-1}\left\{(\hat c[u_3] - \hat{\Delta}[v])D_2 - \hat c[u_2]D_3\right\}, & W^{-1}\Delta \\
   \displaystyle \frac{v_2}{\Delta}\hat\Delta W^{-1}\left\{(\hat c[u_3] - \hat{\Delta}[v])D_2 - \Delta\hat{c}{\hat\Delta}^{-1}D_3\right\}
+ \frac{v_3}{\Delta}D_2, &\displaystyle \frac{v_2}{\Delta}\hat\Delta W^{-1}\Delta
\end{array}\right)\nonumber\\
\label{R-1II}
\end{eqnarray}
where $W = \Delta\hat c - \hat c[u_2]\hat\Delta$. Using its adjoint in the formula \eqref{R^+^-1II} we obtain the null result $H_0 = 0$.

Hence we need the next Hamiltonian density $H_2$ in the hierarchy of the system $II$. We apply the relation \eqref{delH2I} to obtain
\begin{equation}
 H_2 = v\hat c[u]\Delta = v(c_7u_3+c_8u_1)(a_8u_{12}+a_{10}u_{22}+a_{11}u_{23})
 \label{H2II}
\end{equation}
with the variational derivatives
\[\delta_uH_2 = \hat\Delta\bigl[v_2\hat c[u]\bigr] + \hat\Delta[v]\hat c[u_2] - \Delta\hat c[v],\qquad \delta_vH_2 = \Delta\hat c[u ].\]
We also need the Hamiltonian operator $J_{-1} = J_0(R^{-1})^\dagger$ with the result
\begin{equation}
 J_{-1} = \left(\begin{array}{cc}
  - W^{-1},       & W^{-1}\hat\Delta\displaystyle\frac{v_2}{\Delta} \\
  - \displaystyle\frac{v_2}{\Delta}\hat\Delta W^{-1}, & \displaystyle\frac{1}{\Delta}(v_2\hat\Delta W^{-1}\hat\Delta v_2 + v_3D_2 - v_2D_3)\frac{1}{\Delta}
\end{array}\right)
  \label{J-1II}
\end{equation}
which is manifestly skew-symmetric.

Then we can easily check the validity of bi-Hamiltonian representation for the system $II$ in the form
\begin{equation}
\left(
\begin{array}{c}
 u_t\\
 v_t
\end{array}
\right) = J_0
\left(
\begin{array}{c}
 \delta_u H_1\\
 \delta_v H_1
\end{array}
\right) = J_{-1}
\left(
\begin{array}{c}
 \delta_u H_2\\
 \delta_v H_2
\end{array}
\right).
  \label{bi_HamII}
\end{equation}

The first nonlocal flow is obtained by
\begin{equation}
\left(
\begin{array}{c}
 u_{\tau_3} \\
 v_{\tau_3}
\end{array}
 \right) =
 J_1
\left(\begin{array}{c}
  \delta_u H_2 \\
  \delta_v H_2
\end{array}
\right)
 \label{tau3}
\end{equation}
with the explicit form
\begin{eqnarray}
&& u_{\tau_3} = L_{23(t)}^{-1}\left\{\Delta\hat c[v] - v_2\hat\Delta\bigl[\hat c[u]\bigr]\right\} \nonumber\\
&& v_{\tau_3} = \frac{q}{v_2} D_2L_{23(t)}^{-1}\left\{\Delta\hat c[v] - v_2\hat\Delta\bigl[\hat c[u]\bigr]\right\}
+ {\hat c}^2[u] - \frac{\hat c[u_2]\hat c[v]}{v_2}.
 \label{flow3II}
\end{eqnarray}

We must show that this flow commutes with our original flow $u_t=v$, $v_t=q$ of the system $II$. The straightforward check of this fact is impossible because of the nonlocal operator in \eqref{flow3II}. Therefore, we apply the following more sophisticated procedure of checking the compatibility of the system $II$ with the symmetry flow \eqref{flow3II}. We rewrite the equations \eqref{flow3II} in the local form
\begin{equation}
  L_{23(t)}[u_{\tau_3}] = \Delta\hat c[v] - v_2\hat\Delta\bigl[\hat c[u]\bigr],\quad v_{\tau_3} = \frac{q}{v_2}D_2 u_{\tau_3}
+ {\hat c}^2[u] - \frac{\hat c[u_2]\hat c[v]}{v_2}.
 \label{flowII}
\end{equation}
We solve the first equation in \eqref{flowII} with respect to $u_{\tau_3 z_3}$, differentiate this equation with respect to $t$ and substitute $u_t=v$ and $v_t=q$. We will also need $D_2[v_{\tau_3}]$ and $D_3[v_{\tau_3}]$ which we obtain by differentiating the second equation in \eqref{flowII}. After some tedious calculation we find that the first equation in \eqref{flowII} differentiated w.r.t. $t$ with $u_t=v, v_t=q$ determined by system $II$ is identically satisfied as a consequence of equations \eqref{flow3II}, which proves the compatibility of the system $II$ and the flow \eqref{flow3II}.

The invariant solution with respect to this flow is determined by the condition $u_{\tau_3} = 0$, which implies $ v_{\tau_3} = 0$ which, due to \eqref{flowII} implies the equations
\begin{equation}
 \Delta\hat c[v] - v_2\hat\Delta\bigl[\hat c[u]\bigr] = 0,\quad \hat c[u_2]\hat c[v] - v_2{\hat c}^2[u] = 0
 \label{statflowII}
\end{equation}
or, equivalently,
\begin{equation}
  \hat c[u_2]\hat c\bigl[\hat\Delta[u]\bigr] - \Delta{\hat c}^2[u] = 0,\quad \Delta\hat c[v] - v_2\hat\Delta\bigl[\hat c[u]\bigr] = 0.
 \label{statflowIIa}
\end{equation}
Here we have to solve the first equation to determine $u$ and then the second equation to determine $v$ or, alternatively, we can just use $v=u_t$ with the same result.
No symmetry reduction in the number of independent variables needs to occur because of the nonlocality of the flow.

One may wonder if we could obtain another independent equation for $v$ by differentiating the first equation \eqref{statflowIIa} w.r.t. $t$ (which would be bad)
\begin{equation}
\hat c[v_2]\hat c\bigl[\hat\Delta[u]\bigr] + \hat c[u_2]\hat c\bigl[\hat\Delta[v]\bigr]  - \hat\Delta[v_2]{\hat c}^2[u] - \Delta{\hat c}^2[v] = 0.
 \label{D_tstatflowII}
\end{equation}
We apply the operator $\hat\Delta$ to the second equation in \eqref{statflowII} and combine the resulting equation with \eqref{D_tstatflowII} to obtain
\[\hat c\left[\Delta\hat c[v] - v_2\hat\Delta\bigl[\hat c[u]\bigr]\right] = 0\]
which is identically satisfied due to the second equation in \eqref{statflowIIa}. Thus, $t$-differentiation of the first equation \eqref{statflowIIa} does not yield an independent equation which is again a confirmation of the compatibility of the flow \eqref{flow3II} with the system $II$.

\subsection{Hierarchy of system $III$}
\label{III.3}
\setcounter{equation}{0}

According to \eqref{6.11}, system $III$ has the form
\begin{eqnarray}
&& u_t = v
 \label{sysIII}\\
&& v_t = q =\frac{1}{u_{33}}\left\{v_3^2 - c_5L_{23(2)}[v] - c_6L_{13(3)}[v] - c_7L_{23(3)}[v] - c_8L_{23(1)}[v]\right\} .
 \nonumber
\end{eqnarray}
Recursion operator for the system \eqref{sysIII} was obtained in \cite{S_Y}. We present it here in a more compact form by using the relation
$L_{ij(k)}[v]D_j + u_{jk}L_{ij(t)} = v_jL_{ij(k)}$ (with $u_t=v$) and definitions
\[W = c_5L_{23(3)} + c_8L_{13(3)},\quad L = c_5L_{23(2)} + c_6L_{13(3)} + c_7L_{23(3)} + c_8L_{23(1)}.\]
The recursion operator takes the form
\begin{equation}
R = \left(\begin{array}{cc}
 W^{-1}(v_3D_3 - L),            &   -  W^{-1}u_{33} \\
\displaystyle\frac{1}{u_{33}}\left\{v_3D_3W^{-1}(v_3D_3 - L) - L_{23(t)}\right\}, & \displaystyle - \frac{v_3}{u_{33}}D_3W^{-1}u_{33}
\end{array}\right).
 \label{RIII}
\end{equation}
The first Hamiltonian operator \cite{S_Y}
\begin{equation}
 J_0 = \frac{1}{u_{33}}\left(\begin{array}{cc}
  0 , & 1 \\
  -1, & \displaystyle (v_3D_3 + D_3v_3 - L)\frac{1}{u_{33}}
\end{array}\right)
 \label{J0III}
\end{equation}
together with the corresponding Hamiltonian density
\begin{equation}
 H_1 = \frac{1}{2} v^2u_{33}
 \label{H1III}
\end{equation}
yields the first Hamiltonian form of the system $III$ in \eqref{be_Ham}. The second Hamiltonian operator $J_1 = RJ_0$ is obtained by composing the recursion operator \eqref{RIII} with the Hamiltonian operator \eqref{J0III} with the result
\begin{equation}
  J_1 = \left(
\begin{array}{cc}
 W^{-1} , &\displaystyle - W^{-1}D_3\frac{v_3}{u_{33}} \\
\displaystyle \frac{v_3}{u_{33}}D_3W^{-1} , & \displaystyle - \frac{1}{u_{33}}\left(v_3D_3W^{-1}D_3v_3 + L_{23(t)}\right)\frac{1}{u_{33}}
\end{array}\right)
\label{J1III}
\end{equation}
which is manifestly skew-symmetric.
For the corresponding Hamiltonian density $H_0$ we assume the simplest possible ansatz of its dependence only on $v$ but not on derivatives of $v$ which ends up with the expression
\begin{equation}
 H_0 = \left[k(t,z_1)v^2 - (c_8u_1 + c_5u_2 + c_7u_3)v\right]u_{33}
 \label{H0III}
\end{equation}
together with the \textit{existence condition $c_6 = 0$} for such a density. We should note that the second Hamiltonian operator \eqref{J1III} is valid with no such extra conditions and probably it admits the corresponding Hamiltonian density more general than \eqref{H0III} with no additional restrictions. The same remark applies also for the existence condition $c_8c_{10} = c_5c_9$ of density $H_0$ for system $I$ in \eqref{H0I}.
With the condition $c_6 = 0$, the obtained expressions for $J_0$, $H_1$, $J_1$ and $H_0$ yield the bi-Hamiltonian representation \eqref{be_Ham} for the system $III$.

The next Hamiltonian density $H_2$ in the hierarchy of the system $III$ should be generated by the formal adjoint of the recursion operator
\begin{equation}
 \left(\begin{array}{c}
 \delta_uH_2 \\
 \delta_vH_2
\end{array}\right) = R^\dagger
\left(\begin{array}{c}
 \delta_uH_1 \\
 \delta_vH_1
\end{array}\right) =
\left(\begin{array}{c}
 0 \\
 0
\end{array}\right)
 \label{H2III}
\end{equation}
but the null result implies that $H_2 = 0$ and so are all the next members of the hierarchy in the right direction.

Hence, to obtain nontrivial results we need an inverse recursion operator $R^{-1}$ determined by the formulas \eqref{R^-1} to be able to move in the left direction along the hierarchy chain.
We again use more simple formulas given above in \eqref{R^-1simple} with the result
\begin{eqnarray}
&&  R^{-1} =
  \label{R^-1III}\\
&& \left(\begin{array}{cc}
  L_{23(t)}^{-1}v_3D_3,    &    - L_{23(t)}^{-1}u_{33} \\
\displaystyle \frac{1}{u_{33}}\left\{(v_3D_3 - L)L_{23(t)}^{-1}v_3D_3 - W\right\}, & \displaystyle - \frac{1}{u_{33}} (v_3D_3 - L)L_{23(t)}^{-1}u_{33}
\end{array}\right)\nonumber.
\end{eqnarray}
We use the operator adjoint to $R^{-1}$ to generate the Hamiltonian operator $J_{-1}$ moving in the left direction along the hierarchy chain
\begin{eqnarray}
&&\hspace*{-12pt}  J_{-1} = J_0(R^{-1})^\dagger =
  \label{J-1III}\\
&&\hspace*{-12pt} \left(\begin{array}{cc}
  L_{23(t)}^{-1},    & \displaystyle   - L_{23(t)}^{-1}(D_3v_3 - L)\frac{1}{u_{33}} \\
\displaystyle \frac{1}{u_{33}}(v_3D_3 - L)L_{23(t)}^{-1}, & \displaystyle \frac{1}{u_{33}}\!\left\{W - (v_3D_3 - L)L_{23(t)}^{-1}(D_3v_3 - L)\right\}\! \frac{1}{u_{33}}
\end{array}\right)\nonumber.
\end{eqnarray}
Thus, $J_{-1}$ is manifestly skew-symmetric.
The first nonlocal flow is generated by the formula
\begin{equation}
 \left(\begin{array}{c}
 u_{\tau_{-1}} \\
 v_{\tau_{-1}}
\end{array}\right) = J_{-1}
   \left(\begin{array}{c}
 \delta_u H_0 \\
 \delta_v H_0
\end{array}\right)
 \label{tau-1III}
\end{equation}
with the explicit result for the flow \eqref{tau-1III}
\begin{eqnarray}
&& u_{\tau_{-1}} = L_{23(t)}^{-1}(W + 2kL)[v]
\label{nonlocalIII}\\
&& v_{\tau_{-1}} = \frac{1}{u_{33}}\left\{(v_3D_3 - L)L_{23(t)}^{-1}(W + 2kL)[v] - W[2kv - c_5u_2 - c_8u_1]\right\}\nonumber
\end{eqnarray}
where the identities $L[c_5u_2 + c_7u_3 + c_8u_1] = 0$ and $W[u_3] = 0$ have been used. The first equation in \eqref{nonlocalIII} can be written in the local form
\begin{equation}
  L_{23(t)}[u_{\tau_{-1}}] = (W + 2kL)[v]
 \label{nonlocal1III}
\end{equation}
while the second equation becomes
\begin{equation}
  v_{\tau_{-1}} = \frac{1}{u_{33}}\left\{(v_3D_3 - L)u_{\tau_{-1}} - W[2kv - c_5u_2 - c_8u_1]\right\}
 \label{nonlocIII}
\end{equation}
which is more convenient for studying the commutativity of the flow \eqref{nonlocalIII} with the system $III$, similar to the way it was done at the end of subsection \ref{II.3}.

For the stationary flow $u_{\tau_{-1}} = 0$, $v_{\tau_{-1}} = 0$  we have
$W[v] + 2kL[v] = 0$ and the second equation in \eqref{nonlocalIII} becomes $W[2kv - c_5u_2 - c_8u_1]=0$. Explicit solutions to these equations need further analysis and will be published elsewhere.

\subsection{Hierarchy of system $IV$}
\label{IV.3}
\setcounter{equation}{0}

System $IV$ can be written in the compact form
\begin{equation}
 u_t = v,\quad v_t = q = \frac{1}{\Delta}\left\{v_1(\hat\Delta[v] - \hat c[u_2]) + v_2\hat c[u_1]\right\}
 \label{IV}
\end{equation}
where
\begin{equation}
 \hat\Delta = a_7D_1 + a_8D_2 + a_9D_3,\quad \hat c = c_1D_1 + c_3D_2 + c_4D_3
 \label{hatoper}
\end{equation}
and $\Delta=\hat\Delta[u_1]$. The notation is similar to the one for the system $II$ but the definitions of $\hat\Delta$ and $\hat c$ are different and the two systems are also completely different, the system $II$ depending on five parameters while system $IV$ depends on six parameters.

Recursion operator for system $IV$, as obtained in \cite{S_Y}, has the form
\begin{equation}
 R = \left(\begin{array}{cc}
 L_{12(t)}^{-1}v_1\hat\Delta , & - L_{12(t)}^{-1}\Delta \\
 \displaystyle\frac{q}{v_1}D_1L_{12(t)}^{-1}v_1\hat\Delta - \hat c, & \displaystyle\frac{1}{v_1}\left(\hat c[u_1] - qD_1L_{12(t)}^{-1}\Delta\right)
\end{array}\right)
 \label{RIV}
\end{equation}
where $L_{12(t)} = v_2D_1-v_1D_2$ which implies $L_{12(t)}[v] \equiv 0$. The first Hamiltonian operator has the form
\begin{equation}
 J_0 = \frac{1}{\Delta}\left(\begin{array}{cc}
 0   & 1 \\
 - 1 & K_{11}\frac{1}{\Delta}
\end{array}\right)
 \label{J0IV}
\end{equation}
where $K_{11} = v_1\hat\Delta + D_1\hat\Delta[v] + \hat c[u_1]D_2 - \hat c[u_2]D_1$, with the corresponding Hamiltonian density
\begin{equation}
 H_1 = \frac{v^2}{2} \Delta.
 \label{H1IV}
\end{equation}
The second Hamiltonian operator $J_1 = RJ_0$ reads \cite{S_Y}
\begin{equation}
 J_1 =\left(\begin{array}{cc}
  L_{12(t)}^{-1},  &\displaystyle - \left(L_{12(t)}^{-1}D_1q - \frac{\hat{c}[u_1]}{\Delta}\right)\frac{1}{v_1} \\[2pt]
  \displaystyle\frac{1}{v_1}\left(qD_1L_{12(t)}^{-1} - \frac{\hat{c}[u_1]}{\Delta}\right), & J_1^{22}
 \end{array}
 \right)
\label{J_1IV}
\end{equation}
where
\begin{eqnarray}
&& J_1^{22} = - \hat{c}\frac{1}{\Delta} + \frac{\hat{c}[u_1]}{\Delta}\hat{\Delta}\frac{1}{\Delta} - \frac{q}{v_1}D_1L_{12(t)}^{-1}D_1\frac{q}{v_1}
+ \frac{q}{v_1}D_1\frac{\hat{c}[u_1]}{v_1\Delta}\nonumber\\
&&\mbox{} +  \frac{\hat{c}[u_1]}{\Delta v_1}D_1\frac{q}{v_1} - \frac{\hat{c}[u_1]}{\Delta v_1}L_{12(t)}\frac{\hat{c}[u_1]}{v_1\Delta}.
\label{J1_22IV}
\end{eqnarray}
Formulas \eqref{J_1IV} and \eqref{J1_22IV} show that $J_1$ is manifestly skew-symmetric: \\ $J_1^\dagger = - J_1$.

However, we encounter difficulties, similar to the ones for the system $II$, in finding the Hamiltonian density $H_0$ corresponding to $J_1$ in the bi-Hamiltonian representation \eqref{be_Ham} of system $IV$. They are related to the fact that $v$ belongs to the kernel of the operator $L_{12(t)}$, so that to enforce the relation $L_{12(t)}^{-1}L_{12(t)} = 1$ we had to skip $v$ which is needed to reproduce the correct second equation in \eqref{be_Ham}. Therefore, to determine the correct $H_0$ we apply the relation
\begin{equation}
 \left(\begin{array}{c}
 \delta_u H_0 \\
 \delta_v H_0
\end{array}\right) = (R^\dagger)^{-1}
\left(\begin{array}{c}
 \delta_u H_1 \\
 \delta_v H_1
\end{array}\right)
 \label{R^+^-1}
\end{equation}
inverse to \eqref{R^+}. Hence at this point we again need an adjoint inverse recursion operator. Let the recursion operator \eqref{RIV} and its inverse be written in the form
\[R = \left(\begin{array}{cc}
  a & b \\
  c & d
\end{array}\right),\quad
R^{-1} = \left(\begin{array}{cc}
  e & f \\
  g & h
\end{array}\right).\]
Then $R^{-1}$ is determined by the relations \eqref{R^-1simple}
\begin{eqnarray}
&&\hspace*{-19pt} f = \left(\hat c[u_1]\hat\Delta - \Delta \hat c\right)^{-1}\!\Delta,\quad e = - fdb^{-1}\nonumber\\
&&\hspace*{-19pt} = \left(\hat{c}[u_1]\hat\Delta - \Delta \hat c\right)^{-1}\left\{\frac{\hat c[u_1]}{v_1}L_{12(t)}
- \left(\hat\Delta[v] + \frac{1}{v_1}L_{12(t)}[\hat c[u]]\right)D_1\right\} \nonumber\\
&&\hspace*{-19pt} h = - b^{-1}a f = \frac{v_1}{\Delta}\hat\Delta \left(\hat c[u_1]\hat\Delta - \Delta \hat c\right)^{-1}\Delta \nonumber\\
&&\hspace*{-19pt} g = - h c a^{-1}
\label{R^-1IV} \\
&&\hspace*{-19pt} = - \frac{v_1}{\Delta}\hat\Delta\left(\hat c[u_1]\hat\Delta - \Delta \hat c\right)^{-1}
\left\{\left(\hat\Delta[v] + \frac{1}{v_1}L_{12(t)}[\hat c[u]]\right)D_1 - \Delta\hat c{\hat\Delta}^{-1}\frac{1}{v_1}L_{12(t)}\right\}
\nonumber
\end{eqnarray}
and its adjoint has the form
\[(R^{-1})^\dagger = (R^\dagger)^{-1} = \left(\begin{array}{cc}
  e^\dagger & g^\dagger \\
  f^\dagger & h^\dagger
\end{array}\right)\]
where
\begin{eqnarray}
&& e^\dagger = \left\{L_{12(t)}\frac{\hat c[u_1]}{v_1} - D_1\left(\hat\Delta[v] + \frac{1}{v_1}L_{12(t)}[\hat c[u]]\right)\right\}W^{-1}
\nonumber\\
&& g^\dagger = \left\{D_1\left(\hat\Delta[v] + L_{12(t)}[\hat c[u]]\frac{1}{v_1}\right) - L_{12(t)}\frac{1}{v_1}{\hat\Delta}^{-1}\hat c\Delta \right\}
W^{-1}\hat\Delta\frac{v_1}{\Delta} \nonumber\\
&& f^\dagger = - \Delta W^{-1}, \quad h^\dagger = \Delta W^{-1}\hat\Delta\frac{v_1}{\Delta}
\end{eqnarray}
where $W = \hat\Delta\hat c[u_1] - \hat c\Delta$ and we keep in mind that the square brackets denote the value of an operator.

Using the formula \eqref{R^+^-1} with $\delta_u H_1 = \hat\Delta[vv_1]$, $\delta_v H_1 = v\Delta$, we obtain the result $\delta_u H_0 = 0, \delta_v H_0 = 0$
and $H_0 = 0$. Hence the bi-Hamiltonian representation of the system $IV$ in the form \eqref{be_Ham} is not valid.
Therefore, we search for a satisfactory Hamiltonian density moving in opposite direction from $H_1$ to $H_2$ via the relation
\begin{equation}
 \left(\begin{array}{c}
 \delta_u H_2 \\
 \delta_v H_2
\end{array}\right) = R^\dagger
\left(\begin{array}{c}
 \delta_u H_1 \\
 \delta_v H_1
\end{array}\right).
 \label{R^+H2}
\end{equation}
The result is
\begin{equation}
 H_2 = - v\hat c[u]\Delta
 \label{H2IV}
\end{equation}
with the variational derivatives
\[\delta_uH_2 = \hat c[v]\Delta - \hat\Delta[v]\hat c[u_1] - \hat\Delta\bigl[v_1\hat c[u]\bigr],\quad \delta_vH_2 = - \Delta\hat c[u]. \]
As we will immediately see, $H_2$ corresponds naturally to the Hamiltonian operator $J_{-1} = J_0(R^{-1})^\dagger$ with the explicit expression
\begin{equation}
 J_{-1} = \left(\begin{array}{cc}
 - W^{-1}, &\displaystyle W^{-1}\hat\Delta\frac{v_1}{\Delta}\\
\displaystyle - \frac{v_1}{\Delta}\hat\Delta W^{-1}, &\displaystyle \frac{1}{\Delta}\left(v_1\hat\Delta W^{-1}\hat\Delta v_1 - L_{12(t)}\right)\frac{1}{\Delta}
\end{array}\right)
 \label{J_- 1IV}
\end{equation}
which is manifestly skew-symmetric. Now, a straightforward check proves the validity of the following bi-Hamiltonian representation of the system $IV$
\begin{equation}
\left(
\begin{array}{c}
 u_t\\
 v_t
\end{array}
\right) = J_0
\left(
\begin{array}{c}
 \delta_u H_1\\
 \delta_v H_1
\end{array}
\right) = J_{-1}
\left(
\begin{array}{c}
 \delta_u H_2\\
 \delta_v H_2
\end{array}
\right).
  \label{bi_HamIV}
\end{equation}

To discover higher (nonlocal) flows, we consider
\begin{equation}
  \left(
\begin{array}{c}
 u_{\tau_3}\\
 v_{\tau_3}
\end{array}
\right) =
J_1
\left(
\begin{array}{c}
 \delta_u H_2\\
 \delta_v H_2
\end{array}
\right).
 \label{tau3IV}
\end{equation}
The explicit form of the flow \eqref{tau3IV} reads
\begin{eqnarray}
&& u_{\tau_3} = L_{12(t)}^{-1}\left\{\Delta\hat c[v] - v_1\hat\Delta\bigl[\hat c[u]\bigr]\right\} \nonumber\\
&& v_{\tau_3} = \frac{1}{\Delta}\left(\hat\Delta[v] + \frac{1}{v_1}L_{12(t)}\bigl[\hat c[u]\bigr]\right)D_1L_{12(t)}^{-1}
\left\{\Delta\hat c[v] - v_1\hat\Delta\bigl[\hat c[u]\bigr]\right\} \nonumber\\
&&\mbox{} - \frac{\hat c[u_1]\hat c[v]}{v_1} + {\hat c}^2[u].
\label{nonlocIY}
\end{eqnarray}
The second equation \eqref{nonlocIY} can be rewritten as
\begin{equation}
  v_{\tau_3} = \frac{1}{\Delta}\left(\hat\Delta[v] + \frac{1}{v_1}L_{12(t)}\bigl[\hat c[u]\bigr]\right)D_1 u_{\tau_3}
  - \frac{\hat c[u_1]\hat c[v]}{v_1} + {\hat c}^2[u].
\label{nonlocaIV}
\end{equation}
Commutativity of system $IV$ flow and nonlocal symmetry flow \eqref{nonlocIY} can be proved by the procedure similar to the one presented at the end of the subsection \ref{II.3}.

Stationary solutions $u_{\tau_3} = 0$, $v_{\tau_3} = 0$ of the flow \eqref{nonlocIY} are determined by the equations
\begin{eqnarray}
&& \Delta\hat c[v] - v_1\hat\Delta\bigl[\hat c[u]\bigr] = 0\nonumber\\
&& \hat c[u_1]\hat c[v] - v_1{\hat c}^2[u] = 0.
 \label{statIV}
\end{eqnarray}
Solutions of these equations will be published elsewhere. They will not experience symmetry reduction in the number of independent variables because of nonlocality of the flow.

\section{Conclusion}

We have carried out a detailed analysis of our four new bi-Hamiltonian systems in 3+1 dimensions.
Point symmetries and conserved densities generating these symmetries have been presented. Hierarchies of these four systems were studied showing the important role played by the inverse recursion operators $R^{-1}$. For systems $II$ and $IV$ such operators are necessary to obtain a correct bi-Hamiltonian representation of the system, while for systems $I$ and $III$ operators $R^{-1}$ are utilized to discover nonlocal symmetry flows. We have explicitly constructed first nonlocal symmetry flows in the hierarchy for each of the four heavenly systems. Stationary solutions of the latter flows do not need to admit symmetry reduction in the number of independent variables and therefore the corresponding (anti-)self-dual gravitational metrics will not admit Killing vectors, which is a characteristic feature of the $K3$ gravitational instanton. Explicit form of solutions invariant w.r.t. nonlocal symmetry flows is now in progress. The description of (anti-)self-dual gravity governed by our new bi-Hamiltonian heavenly systems will be published elsewhere.

\section*{Acknowledgments}

This work was supported by Research Fund of the Y{\i}ld{\i}z Technical University. Project Number: 3462.



%% References


\bibliographystyle{elsarticle-num}
\bibliography{<your-bib-database>}



\end{document}